%% file: MainSequenceFIRE.tex
\documentclass[useAMS,usenatbib]{mn2e}

\usepackage{lscape}
\usepackage{tabularx}
\usepackage{xtab}
\usepackage{xspace}
\usepackage{graphicx}
\usepackage{placeins}
\usepackage{amsfonts,amsmath,amssymb}
\usepackage{url}
\usepackage[usenames,dvipsnames]{color}
\usepackage{times}

\usepackage{enumerate}
\include{JournalAbbr}

\newcommand{\msun}{~\mathrm{M}_{\odot}}
\newcommand{\mstar}{~M_{\star}}
\newcommand{\sfrten}{SFR$_\text{10 Myr}$\xspace}
\newcommand{\sfrth}{SFR$_\text{200 Myr}$\xspace}

\title[Observational signatures of bursty star formation in galaxies]{(Star)bursts of FIRE: observational signatures of bursty star formation in galaxies}
\author[Sparre et al.]{\parbox[t]{\textwidth}{
		Martin Sparre$^{1,2}$\thanks{E-mail:sparre@dark-cosmology.dk}\thanks{Sapere Aude Fellow},
		Christopher C. Hayward$^{3,4,5}$,
		Robert~Feldmann$^6$\thanks{Hubble Fellow},
		Claude-Andr\'{e}~Faucher-Gigu\`{e}re$^7$,
		Alexander~L.~Muratov$^8$,
		Du\v{s}an~Kere\v{s}$^8$,		
		Philip~F.~Hopkins$^4$		
		\vspace*{6pt}} \\
$^1$Heidelberger Institut f{\"u}r Theoretische Studien, Schloss-Wolfsbrunnenweg 35, 69118 Heidelberg, Germany\\
$^2$Dark Cosmology Centre, Niels Bohr Institute, University of Copenhagen, Juliane Maries Vej 30, 2100 Copenhagen, Denmark\\
$^3$Center for Computational Astronomy, Flatiron Institute, 162 5th Avenue, New York, NY 10010, USA\\
$^4$TAPIR 350-17, California Institute of Technology, 1200 E. California Boulevard, Pasadena, CA 91125, USA\\
$^5$Harvard--Smithsonian Center for Astrophysics, 60 Garden Street, Cambridge, MA 02138, USA\\
$^6$Department of Astronomy, University of California, Berkeley, CA 94720-3411, USA\\
$^7$Department of Physics and Astronomy and CIERA, Northwestern University, 2145 Sheridan Road, Evanston, IL 60208, USA\\
$^8$Department of Physics, Center for Astrophysics and Space Science, University of California at San Diego, 9500 Gilman Drive, La Jolla, CA 92093, USA}

\begin{document}

\date{\today}

\pagerange{\pageref{firstpage}--\pageref{lastpage}} \pubyear{2015}
\maketitle

\label{firstpage}

\begin{abstract}
Galaxy formation models are now able to reproduce observed relations such as the relation between galaxies' star formation rates (SFRs) and stellar masses ($M_*$) and the stellar mass--halo mass relation. We demonstrate that comparisons of the short-timescale variability in galaxy SFRs with observational data provide an additional useful constraint on the physics of galaxy formation feedback. We apply SFR indicators with different sensitivity timescales to galaxies from the Feedback in Realistic Environments (FIRE) simulations. We find that the SFR--$M_*$ relation has a significantly greater scatter when the H$\alpha$-derived SFR is considered compared with when the far-ultraviolet (FUV)-based SFR is used. This difference is a direct consequence of bursty star formation because the FIRE galaxies exhibit order-of-magnitude SFR variations over timescales of a few Myr. We show that the difference in the scatter between the simulated H$\alpha$- and FUV-derived SFR--$M_*$ relations at $z=2$ is consistent with observational constraints. We also find that the H$\alpha$/FUV ratios predicted by the simulations at $z=0$ are similar to those observed for local galaxies except for a population of low-mass ($M_* \lesssim 10^{9.5} \msun$) simulated galaxies with lower H$\alpha$/FUV ratios than observed. We suggest that future cosmological simulations should compare the H$\alpha$/FUV ratios of their galaxies with observations to constrain the feedback models employed.
\end{abstract}

\section{Introduction}

Most massive spiral galaxies in the present-day Universe are in a quasi-equilibrium in which the formation and destruction of giant molecular clouds (GMCs), and the subsequent formation of stars, are regulated by various feedback processes and the infall of gas. However, some galaxies appear to be out of equilibrium in the sense that they are forming stars so rapidly that they will deplete their gas reservoirs on timescales of $10-100$ Myr \citep{2011ApJ...739L..40R,2011ApJ...743..121A,2014ApJ...789...96A,2014ApJ...791...17M}. Such galaxies are referred to as \emph{starburst} galaxies.\footnote{It is worth noting that the term `starburst' is an ambiguous concept that is used in many ways in the literature; see \citet{2009ApJ...698.1437K} for a thorough discussion. For this reason, one should use caution when comparing our results to the literature, especially that of the high-redshift galaxy community, in which the term `starburst' is often used to mean `a galaxy with a high SFR'.}
The definition of a starburst often involves the concept of a timescale -- typically either the gas consumption timescale ($M_\text{gas}/$SFR) or the stellar mass doubling timescale ($M_*/$SFR) -- that is short compared to the lifetime of the galaxy \citep{2009ApJ...698.1437K}. An alternative definition of a starburst relies on comparing SFR indicators that are sensitive to different timescales. For example, the H$\alpha$ nebular emission line and UV continuum fluxes typically trace a galaxy's SFR averaged over the last $\simeq 10$ and $\simeq 200$ Myr, respectively \citep{2012ARA&A..50..531K,2013seg..book..419C}. 
Thus, if a galaxy has an increased H$\alpha$/UV flux ratio compared to the overall population of galaxies, this galaxy may have had a short burst of star formation within the last 10 Myr. Observations of H$\alpha$- and UV-derived SFRs indeed show that short ($\simeq 10$ Myr) bursts play an important role in local dwarf galaxies \citep{2012ApJ...744...44W}. A similar conclusion was reached in a study of galaxies from the Sloan Digital Sky Survey \citep{2014MNRAS.441.2717K}. Both studies found a starburst fraction that decreases with increasing stellar mass. Comparisons of the H$\alpha$- and UV-derived SFRs have also been used to constrain the role of bursty star formation at high redshift \citep{2015arXiv151108808S}.

Understanding the time variability of galaxy star formation histories is critical for many observational reasons. For example, it has been shown that bursty star formation can potentially bias high-redshift galaxy surveys because galaxies in an active burst state will be preferentially selected \citep{2014arXiv1408.5788D}. Also, spectral energy distribution modeling is routinely used to infer physical properties of galaxies (see \citealt{Walcher2011} and \citealt{Conroy2013} for recent reviews). Because the results of such modeling can be quite sensitive to the star formation histories used to generate the model library \citep[e.g.,][]{Pacifici2013,Pacifici2015,Michalowski2014,Simha2014,SH2015}, it is desirable that the input star formation histories are as physically motivated as possible.

Large-volume cosmological simulations of galaxy formation must rely on `sub-resolution' models \citep{2003MNRAS.339..289S,2006MNRAS.373.1265O,2012MNRAS.423.1726S,2012MNRAS.426..140D} because the sub-kiloparsec structure of the ISM is not resolved. Instead, stars are stochastically formed at a rate determined by the local gas density, and a self-regulated ISM is achieved by imposing an effective equation of state that attempts to account for unresolved feedback processes. Examples of state-of-the-art simulations that use such sub-resolution physics models include the Illustris \citep{2014Natur.509..177V}, EAGLE \citep{2014arXiv1407.7040S} and MassiveBlack-II \citep{2015MNRAS.450.1349K} simulations. In contrast, several recent simulations have attempted to physically model the effects of feedback on star-forming clouds instead of imposing a sub-resolution description of them \citep{2006MNRAS.373.1074S,2008Sci...319..174M,2010Natur.463..203G,2012ApJ...761...71Z,2013MNRAS.429.3068T}.
The star formation histories of galaxies simulated with explicit `resolved' feedback are typically more bursty than when a sub-resolution ISM model is used; the latter typically result in a SFR variability timescale of $\gtrsim 100$ Myr \citep{2014arXiv1409.0009S}. However, no galaxy simulations include first-principle calculations of feedback, and there is thus always some uncertainty inherent in the parameterisations or implementations of feedback processes. Consequently, it is a priori unknown whether the very bursty star formation histories of resolved-feedback simulations or the smoother star formation histories of sub-resolution ISM models are more representative of reality. Thus, if feasible, comparisons of the burstiness of simulated and real galaxies' star formation histories may provide an important diagnostic that can inform feedback models.

The \emph{Feedback in Realistic Environments} \citep[FIRE;][]{2013arXiv1311.2073H} project\footnote{The FIRE website is \url{http://fire.northwestern.edu/}.} simulates galaxies in cosmological environments
with a model for explicit stellar feedback. In the FIRE simulations, energy and momentum input from young stars and SN explosions are directly calculated
using a stellar population synthesis model. Feedback operates on the scales of star-forming clouds within the ISM, without artificial ingredients such as
suppressed cooling and hydrodynamical decoupling that are often used in other cosmological simulations. The FIRE simulations are also some of the
highest-resolution cosmological simulations at a given mass that have been performed to date. The physical model for stellar feedback used in the FIRE
simulations has successfully produced star-forming galaxies that obey the Kennicutt--Schmidt relation \citep{1998ApJ...498..541K},
which relates the gas and SFR surface densities of galaxies. The simulations also exhibit good agreement with the stellar mass -- halo mass relation inferred from
abundance matching \citep[see fig. 4 in][]{2013arXiv1311.2073H} and the observed SFR--$M_*$ relation in the local Universe. Given the FIRE model's success in reproducing these observational
constraints, it is natural to consider which tests can further constrain the physical fidelity of the model.

The aim of this paper is to characterise the time variability of star formation in a suite of cosmological `zoom-in' simulations from the FIRE project. By modeling the behaviour of different SFR indicators, we will test whether the burstiness of the FIRE galaxies is consistent with observations of real galaxies. The remainder of this paper is organised as follows. Section~\ref{SFRIndicatorIntro} introduces the different SFR indicators considered, and Section~\ref{SimulationOverview} summarises the details of the simulations used in this work. Section~\ref{SFRMSTAR} studies how the scatter in the SFR--$M_*$ relation probes bursty star formation. Furthermore, we study the signature of bursty star formation cycles in individual galaxies (by comparing the SFR derived by H$\alpha$ and far-UV indicators) at $z=0$. Section~\ref{SNSuite} reveals how the presence of supernova feedback influences the burstiness of galaxies. Section~\ref{Discussion} discusses some implications of our results and Section~\ref{sec:conclusions} summarises the primary conclusions of this work.

\section{Star formation rate indicators}\label{SFRIndicatorIntro}

We consider two theoretical indicators, SFR$_\text{10 Myr}$ and SFR$_\text{200 Myr}$, which correspond to the SFR of a galaxy averaged over the preceding 10 and 200 Myr, respectively. We also consider two observationally motivated SFR indicators: SFR(H$\alpha$), the SFR inferred from the H$\alpha$ recombination line luminosity, and the SFR inferred from the far-ultraviolet (FUV) continuum luminosity, SFR(FUV). We now describe how we
calculate SFR(H$\alpha$) and SFR(FUV) via stellar synthesis modelling.

\subsection{Stellar synthesis modeling with the SLUG code} \label{SLUG}

We use the SLUG code \citep{2012ApJ...745..145D,2014MNRAS.444.3275D,2015arXiv150205408K}, which, given a star formation history, calculates the spectral energy distribution of a galaxy. To calculate the spectrum, we input the star formation history of a galaxy over the past 200 Myr. The H$\alpha$ and FUV fluxes from stellar populations older than 200 Myr are negligible, so it is unnecessary to consider the SFH at longer look-back times. We assume the stellar population to follow solar-metallicity Geneva-tracks with no rotation \citep{2012A&A...537A.146E}. The stellar atmospheres are treated as in {\sc starburst99} \citep{1999ApJS..123....3L,2005ApJ...621..695V,2010ApJS..189..309L,2014ApJS..212...14L}, in which OB star atmospheres are from \citet{2001A&A...375..161P}, Wolf-Rayet star atmospheres are from \citet{1998ApJ...496..407H}, and all other stellar atmospheres are from \citet{1997A&AS..125..229L}. We assume a fully sampled \citet{2001MNRAS.322..231K} initial mass function (IMF). To derive a UV-based SFR, we calculate the flux of the stellar continuum transmitted through the GALEX FUV filter. We assume that the H$\alpha$-derived SFR is proportional to the flux of ionizing photons. With these choices, we can directly compare with H$\alpha$- and FUV-derived SFRs
and H$\alpha$/FUV flux ratios. We make no attempt to model dust attenuation in the present work; thus, our predictions should be compared with dust-corrected observations.

\begin{figure}
\centering
\includegraphics[width = 0.48 \textwidth]{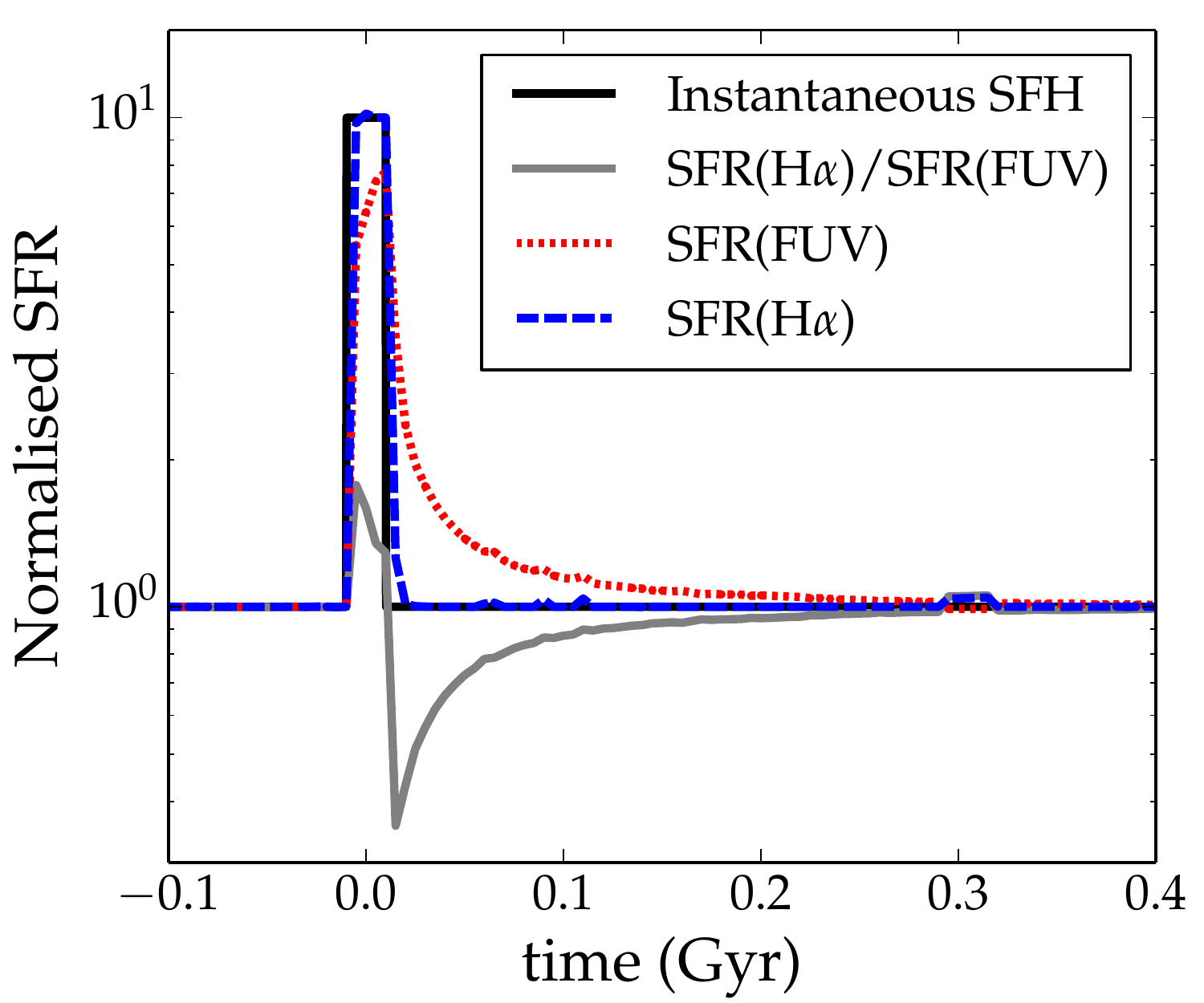}
\caption{The response of the H$\alpha$ (\emph{blue} \emph{dashed}) and FUV (\emph{red} \emph{dotted}) SFR indicators to the
  instantaneous star formation history given by Equation \eqref{SFH_IDEAL}, which is shown by the \emph{black} \emph{solid} line. The ratio between the H$\alpha$- and FUV-derived SFRs (\emph{grey} \emph{solid} line) is sensitive to the presence of short-timescale SFR variability.}
\label{Fig00_SFH_SHUTOFF}
\end{figure}

\subsection{The behaviour of the different SFR indicators}

\begin{table*}
\centering
\begin{tabular}{c|cccc|c|cc|c} 
\hline\hline 
Simulation name  & $m_\text{b}$& $\epsilon_\text{b}$&$m_\text{dm}$&$\epsilon_\text{dm}$&$z$ &$N_\text{gal}$& Fraction &Note \\
&$\msun$ &  pc& $\msun$& pc &&&\% \\
\hline
m10& $2.6\times 10^2$ & 3 & $1.3\times 10^3$ & 30&0.0, 0.2, 0.4&3& 100\\
m11& $7.1\times 10^3$ & 7 & $3.5\times 10^4$ & 70&0.0, 0.2, 0.4&9& 69\\
m12i&$5.0\times 10^4$ & 14 & $2.8\times 10^5$ & 140&0.0, 0.2, 0.4&16&67\\
m12q&$7.1\times 10^3$ & 10 & $2.8\times 10^5$ & 140&0.0, 0.2, 0.4&5&45\\
\hline
m12v&$3.9\times 10^4$ & 10 & $2.0\times 10^5$ & 140&2&-&-& used in Sec. \ref{Modboost}\\
\hline
MassiveFIRE -- HR&$3.3\times 10^4$&9&$1.7\times 10^5$&143&2&786&14\\
\hline
FG15& $5.9\times 10^4$&9&$2.9\times 10^5$&143&2&142&76\\
\hline\hline
\end{tabular}
  \caption{Details of the zoom-in simulations used in this paper. The table includes the baryonic mass resolution ($m_\text{b}$),
  the minimum physical baryonic softening length ($\epsilon_\text{b}$), the dark matter mass resolution ($m_\text{dm}$),
  the minimum physical dark matter softening length ($\epsilon_\text{dm}$), and the redshift at which the simulation is analysed
  in this paper. $N_\text{gal}$ is the total number of H$\alpha$ and FUV measurements in our samples. At $z=2$ $N_\text{gal}$ is the number of different galaxies. To study low-redshift galaxies we have constructed a joint sample of galaxies at $z=0$, $0.2$ and $0.4$. The same galaxy can therefore be included multiple times (but at three different snapshots) in this sample, so here $N_\text{gal}$ does not necessarily corresponds to different galaxies. The galaxies in our sample are selected to have formed more than 50 stellar particles in the previous 200 Myr. In the column \emph{Fraction}, we list the percentage our sample represents of the total number of galaxies that would have been included if this cut had not been performed.}
\label{table:FIREoverview}
\end{table*}

To quantify bursty star formation histories, we study the responses of four different SFR indicators to a rapid change in a galaxy's SFR. We consider a steadily star-forming galaxy in which a 20-Myr-long burst that increases the SFR by a factor of 10 occurs:
\begin{align}
\text{SFR} = 
\begin{cases}
10 \msun \text{yr}^{-1}& \text{if } |t|<10  \text{ Myr,}\\
1 \msun \text{yr}^{-1} & \text{otherwise.}\label{SFH_IDEAL}
\end{cases}
\end{align}
In this illustrative example, we show the effect of a 20-Myr-long burst, which is a typical burst period in our simulations (we later demonstrate this in Figure~\ref{Fig321_MassiveFIRE_1Myr}).

Figure~\ref{Fig00_SFH_SHUTOFF} shows the star formation history specified by Equation \eqref{SFH_IDEAL}
together with H$\alpha$- and FUV-based SFRs and the ratio between them. SFR(H$\alpha$) has a fast response to the change in the instantaneous SFR and is a good indicator of the `instantaneous' SFR of the galaxy. SFR(FUV) has a slower response. The ratio SFR(H$\alpha$)/SFR(FUV) is very sensitive to the burstiness of a galaxy's star formation history.
During the 20-Myr-long burst, this quantity is significantly elevated above the equilibrium
value for a constant SFR (the equilibrium value is 1.0 with our choice of
normalisation), and in the $\simeq$200 Myr after the burst, it is less than
this equilibrium value because SFR(FUV) is still increased by the burst at this time but SFR(H$\alpha$) is not. The statistical distribution of SFR(H$\alpha$)/SFR(FUV) for a sample of galaxies is therefore an efficient way to quantify the importance of short-timescale bursts: galaxies currently undergoing a short-timescale burst have SFR(H$\alpha$)/SFR(FUV)$\,>1$, and galaxies that have experienced a burst at lookback times of 10 Myr
$<t\lesssim$ 200 Myr have SFR(H$\alpha$)/SFR(FUV)$\,<1$. We refer to the latter galaxies as being in a \emph{post-burst} phase.

During the time where SFR(H$\alpha$)$/$SFR(FUV) is affected by the burst, $-$10 Myr $<t\lesssim$ 200 Myr, the median value of this ratio is 0.9. Bursts of star formation will therefore not only increase the scatter in the ratio of two SFR indicators with different timescales but also affect the overall normalisation for a galaxy distribution.

In the remaining parts of this paper, we will use SFR$_\text{10 Myr}$ and SFR$_\text{200 Myr}$ to gain theoretical insight into the SFR variability of the simulated galaxies, and we will use SFR(H$\alpha$) and SFR(FUV) when comparing directly to observations.

\section{Overview of the FIRE simulations}\label{SimulationOverview}

The goal of the FIRE project is to understand how
feedback (thus far only stellar feedback) regulates the formation of
galaxies in the $\Lambda$CDM cosmology.
The code used for the simulations analysed in this work is a heavily modified version of the
{\sc gadget} code \citep{2001NewA....6...79S,2005MNRAS.364.1105S}, {\sc gizmo}
\citep{Hopkins2014,Hopkins2015}.\footnote{A public version of {\sc GIZMO} can be downloaded from
\url{www.tapir.caltech.edu/~phopkins/Site/GIZMO.html}.} The hydrodynamical equations were solved with the pressure-based formulation of the smoothed particle hydrodynamics method \citep{2013MNRAS.428.2840H}.

The simulations are
performed using a multi-scale (`zoom-in') technique in which the resolution is high near the
galaxy of interest and the structure on larger scales is more coarsely resolved.
The dark matter is modeled using collisionless particles.
Gas cools according to a cooling function that includes
contributions from gas in ionized, atomic, and molecular phases. We follow chemical abundances of nine metal species (C, N, O, Ne, Mg, Si, S, Ca and Fe), with enrichment following each source of mass return individually. During the course of the hydrodynamical calculation, ionization balance of all tracked elements is computed using the ultraviolet background model of \citet{2009ApJ...703.1416F}, and we apply an on-the-fly approximation for self-shielding of dense gas. The molecular fraction of dense gas is calculated following \citet{Krumholz2011}.
Stars are formed from molecular gas that has a number density\footnote{A threshold of 5 cm$^{-3}$ is used in the MassiveFIRE runs, and a threshold of 50 cm$^{-3}$ is used in the other runs presented in Table~\ref{table:FIREoverview}.} $n > 5-50$ cm$^{-3}$ and is locally self-gravitating, and an efficiency of 100\% per free-fall time is
assumed; see \citet{2013arXiv1311.2073H} for details.

The star particles that are formed from star-forming gas are treated as single-age stellar
populations, for which a fully sampled Kroupa IMF \citep{2001MNRAS.322..231K} is assumed.
Stellar feedback in the form of radiation pressure, supernovae, stellar winds, photoionization,
and photoelectric heating is included. The inputs for the feedback models (such as stellar
luminosity and supernova rates) are taken directly from {\sc starburst99}
\citep{1999ApJS..123....3L,2005ApJ...621..695V,2010ApJS..189..309L,2014ApJS..212...14L}.
We refer the reader to \citet{HQM11,HQM12} for details regarding and extensive tests of the stellar
feedback models and to \citet{2013arXiv1311.2073H} for details of the stellar feedback model
as implemented in the FIRE simulations.

\subsection{$z = 0$ simulations: The fiducial FIRE simulations}

To study the burstiness of galaxies at low redshift, we use the m10, m11, m12i and m12q runs from \citet{2013arXiv1311.2073H};
see Table~\ref{table:FIREoverview} for details of these simulations. In order to compute e.g. the H$\alpha$-derived SFR accurately,
we need a high number of star particles formed per unit time (this is further described in section~\ref{SampleSelection}),
which means that we can calculate this quantity reliably for only a few halos per simulation snapshot.
When studying the burstiness of galaxies at low redshift, we therefore build a sample of galaxies that contains all the well-sampled
halos from the above simulations from the time snapshots at $z=0$, $z=0.2$ and $z=0.4$. By choosing these three redshifts, it is ensured
that the 200-Myr time intervals from the different snapshots do not overlap, and the galaxies from these snapshots are all from a cosmic
epoch well after the peak of the global SFR density at $z\simeq 2$ \citep{2006ApJ...651..142H, 2013A&A...556A..55I,2013ApJ...770...57B}.
In figure legends we will simply refer to the galaxies at $z=0$, $z=0.2$ and $z=0.4$ as `$z=0$ galaxies'.

\subsection{$z=2$ simulations}

To build a sample of galaxies at $z= 2$, we use the MassiveFIRE simulation suite \citep{2016MNRAS.458L..14F,2016arXiv160107188F} and the set of \emph{z2hXXX} simulations from \citet{2015MNRAS.449..987F}. We will refer to the latter set of simulations as FG15. The simulations in both suites were performed with high mass resolution
and thus were only run until $z=2$ in order to make it possible to run a large sample of simulations.
Our sample of halos from the MassiveFIRE suite consists of 15 high-resolution (HR) cosmological zoom-in simulations of halos with $M_{200}\simeq 2\times 10^{12}-3\times 10^{13}\,\msun$ using the same feedback model as the simulations presented in \citet{2013arXiv1311.2073H}. The HR runs have a significantly increased resolution compared with the $10^{13} \msun$ halo presented in \citet{2013arXiv1311.2073H}, which we do not analyze in this work; see the details presented in Table~\ref{table:FIREoverview}. The FG15 simulated halos have a similar mass resolution as MassiveFIRE but lower halo masses of $1.9\times 10^{11}-1.2\times 10^{12}\msun$.

In Section~\ref{Modboost} we use the m12v simulation (from Table~\ref{table:FIREoverview}) to study the role of supernova feedback. At $z=2$ m12v only contains 3 galaxies that meet our selection criteria, and this number is negligible compared to the 928 galaxies from the FG15 and MassiveFIRE--HR simulations. For this reason we do not include the m12v galaxies in the $z=2$ sample.

\begin{figure*}
\centering
\includegraphics[width = 0.98 \textwidth]{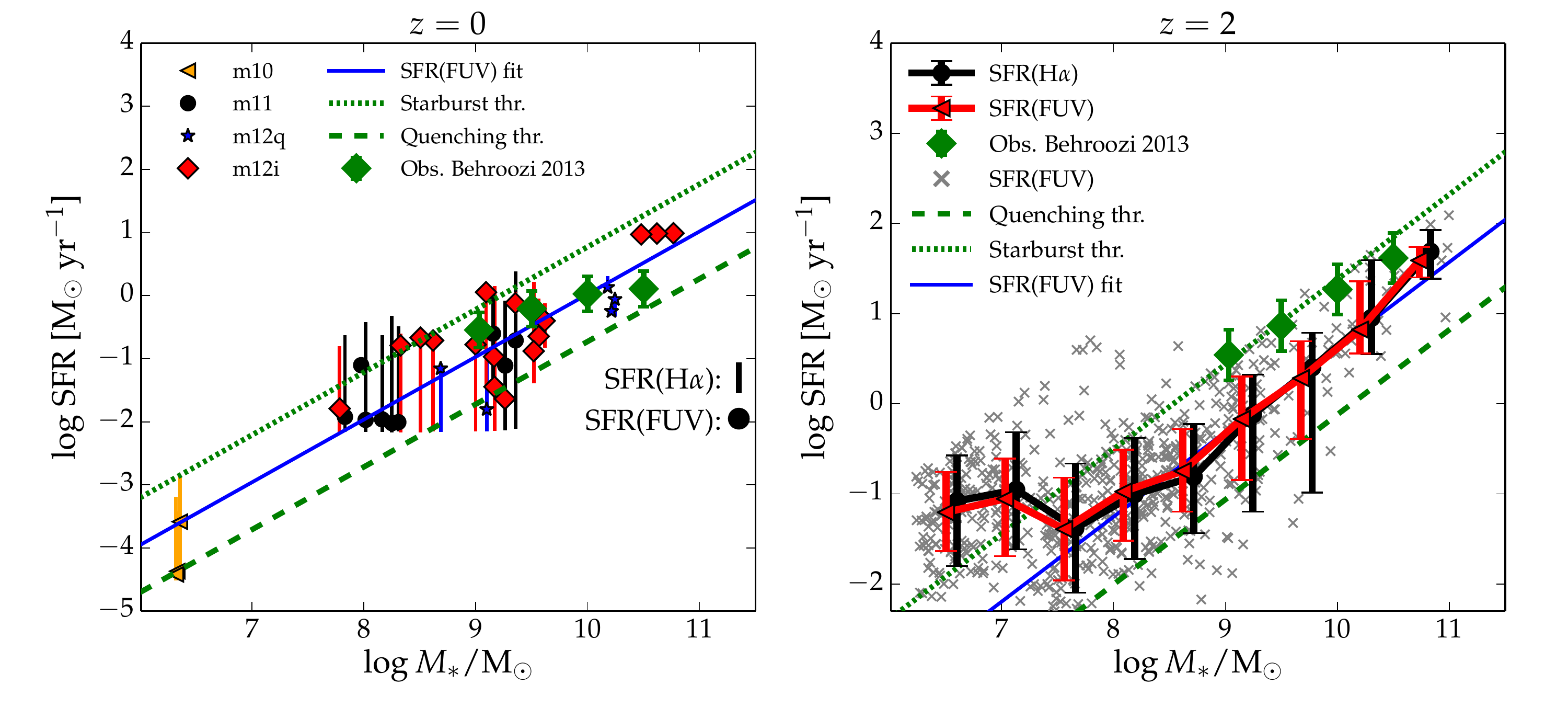}
\caption{\emph{Left panel}: The SFR--$M_*$ relation at $z=0,0.2$ and 0.4 measured using the FUV-derived SFR (different symbols correspond to different simulations, as indicated in the legend), which has a sensitivity timescale of $\simeq 200$ Myr. The H$\alpha$-derived SFR (which has a 10-Myr sensitivity) is evaluated every 10 Myr between lookback times of 0 and 200 Myr, and the interval spanned by the maximum and minimum SFR(H$\alpha$) within this 200-Myr period is denoted by the \emph{vertical error bars}. \emph{Right panel}: The median relations for SFR(H$\alpha$) and SFR(FUV) (\emph{circles} and \emph{triangles}, respectively) at $z=2$; the corresponding error bars show the scatter measured as the 16-84\% percentile in a given mass bin. The $x$ symbols in the background show SFR(FUV) for individual galaxies. The \emph{green dashed lines} and \emph{green dotted lines} indicate the quenching and starburst thresholds, respectively. These are selected to be 0.75 dex away from the fitted relation (\emph{blue line}). The offset of 0.75 dex used to define the thresholds is an observationally motivated choice, that corresponds to galaxies being $2.5\sigma$ outliers assuming a 0.3 dex scatter in the SFR--$M_*$ relation. The \emph{green diamonds} show a compilation of observations from \citet{2013ApJ...770...57B}; the error bars on these points represent the inter-publication variance in the relation, \emph{not the scatter in the relation}. At $z = 0$, the normalisations of the SFR--$M_*$ relations of the simulated and observed galaxies agree well. At $z = 2$, the normalisation of the simulated galaxies' relation is slightly lower than the \citet{2013ApJ...770...57B} compilation. At $z=2$ the SFR--$M_*$ relation for the  $M_*\lesssim 10^{7.5} \msun$ galaxies seems to flatten. This is a selection effect caused by the requirement that 50 star particles have to be formed in the last 200 Myr for a galaxy to be included in our sample (we study these galaxies further in Figure~\ref{Fig06A_BurstSFRFrac}).}
\label{Fig01_SFMS}
\end{figure*}

\subsection{Identification of galaxies and determination of their star formation histories}\label{SampleSelection}

As discussed in Section \ref{SFRIndicatorIntro}, the burstiness of a star formation history
can be quantified by comparing the SFRs averaged over 10 Myr and 200 Myr,
SFR$_\text{10 Myr}$ and SFR$_\text{200 Myr}$, or the observational analogues,
SFR(H$\alpha$) and SFR(FUV). To calculate the star formation histories of the galaxies in
the FIRE simulations, we first identify
halos and subhalos using the AMIGA halo finder
\citep{2004MNRAS.351..399G,2009ApJS..182..608K}. To avoid significant contamination from
low-resolution particles, we require galaxies to have a
mass fraction of high-resolution particles of $f_\text{hires}>0.9$ and containing at least 1000 stellar population particles.
The stellar component of the galaxy is defined as all star particles within 20\% of the virial radius.

We define a sample of galaxies with well-sampled star formation histories, which later is used to study different star formation rate indicators. To build this sample of galaxies we include all galaxies with more than 50 star particles formed within 20\% of the virial radius in the last 200 Myr. This biases the sample towards star-forming galaxies, since it excludes galaxies with SFR$_\text{200 Myr}<50 \times m_* / (\text{200 Myr})$, where $m_*$ is the mean mass of a star particle in a simulation. The fraction of the galaxies that survive this cut is listed in the column \emph{Fraction} in Table~\ref{table:FIREoverview}. The sample includes both central galaxies and satellites.

We then calculate the star formation history over the past 200 Myr based on the age distribution of the stellar particles. When calculating the ratio between two SFR indicators we assume all stellar particles to have identical masses at formation time, and when calculating the actual SFR based on an indicator we correct for the stellar mass loss by assuming that stellar populations have lost 10\% (25\%) of their mass due to stellar evolution in the first 10 Myr (200 Myr) after their formation time \citep[following Fig. 106 in][]{1999ApJS..123....3L}.

When our analysis requires calculation of the H$\alpha$-derived SFR, we assume that the SFR is never less than $m_* / (\text{10 Myr})$ at the time at which we measure the H$\alpha$- and FUV-derived SFRs. This corresponds to the lowest SFR that we can probe given the mass-sampling of our simulations. This requirement is used to avoid galaxies having zero H$\alpha$ flux as a result of the finite mass resolution of the simulations, which would otherwise occur if no stars are formed within the $\simeq$10 Myr prior to a snapshot.

Our low-redshift sample consists of galaxies from snapshots at $z=0, 0.2$ and $0.4$. These galaxies are selected from m10, m11, m12i and m12q. When counting the number of galaxies in this sample ($N_\text{gal}$), the same galaxies might therefore be responsible for several counts because the galaxy is analysed at different snapshots. At $z=2$ the galaxies are from the MassiveFIRE -- HR runs and FG15 simulations. The number of galaxies ($N_\text{gal}$) in the sample from each simulation is quoted in Table~\ref{table:FIREoverview}. Also, the fraction of the galaxies that obeyed our requirement that 50 star particles had to be formed within the last 200 Myr of the simulation is listed for each simulation. The MassiveFIRE simulations have a significantly smaller acceptance fraction than the FG15 runs, potentially because of the differences in terms of how the two sets of halos were selected.

\section{Bursty star formation in the FIRE simulations}\label{SFRMSTAR}

\subsection{The SFR -- $M_*$ relation}\label{subsec_sfrmstar}

Multi-wavelength observations indicate the presence of a correlation between the SFRs and stellar masses, $M_*$, of star-forming galaxies at fixed redshift, and the normalisation of this relation increases with increasing redshift \citep{2004MNRAS.351.1151B,2011A&A...533A.119E,2014MNRAS.443...19R,2015ApJ...801...80L}. The scatter in the relation is roughly mass-independent, with a value of $\simeq 0.2-0.4$ dex \citep{2013ApJ...770...57B,2014ApJS..214...15S}, where the exact value depends on e.g. the sample selection method, the SFR indicator(s) used, and the time evolution of the intrinsic relation within the probed redshift range. This relation can be used to define starbursts galaxies as outliers well above this relation \citep{2011ApJ...739L..40R,2012ApJ...747L..31S}; conversely, galaxies that are well below this relation are referred to as quenched. This relation is also important because its normalisation and scatter may provide important constraints on galaxy formation physics (\citealt{2014MNRAS.438.1985T,2014arXiv1409.0009S,2015MNRAS.450.4486F,2015MNRAS.452.1184M}; but cf. \citealt{2014arXiv1406.5191K}).

In the \emph{left panel} of Figure~\ref{Fig01_SFMS}, we show the SFR--$M_*$ relation for our $z=0,0.2$ and 0.4 simulated galaxies selected according to the sample definition in Section~\ref{SampleSelection}. 
The points indicate the SFR(FUV) values of individual galaxies, and the error bars denote the scatter (measured as the difference between the maximum and minimum of the distribution) of the SFR(H$\alpha$) values of the 20 non-overlapping time bins
spanning the previous 200 Myr. The normalisation agrees well with compilation of observational constraints (from \citealt{2013ApJ...770...57B}, who
compiled the specific SFRs of main sequence galaxies as a function of $M_*$ from various publications; see their Table 5), as already noted by \citet{2013arXiv1311.2073H}. It is clear that some galaxies have very bursty star formation histories because their SFR(H$\alpha$) values can vary by more than an order of magnitude within a 200-Myr period. Another important result is that the SFR(H$\alpha$) variations are much larger in low-mass galaxies ($M_*\lesssim 10^{10}\msun$) than in more-massive galaxies. The \emph{right panel} shows the relation for the $z = 2$ simulated galaxies. The \emph{red triangles} [\emph{black circles}] indicate the median SFR(FUV) [SFR(H$\alpha$)] values in different mass bins, and 16--84\% percentiles of the distributions of the SFR(H$\alpha$) and SFR(FUV) values are denoted by the error bars. The SFR(H$\alpha$) variations are clearly larger than the variations in SFR(FUV); this is a signature of the galaxies' bursty star formation histories.

\begin{figure}
\centering
\includegraphics[width = 0.48 \textwidth]{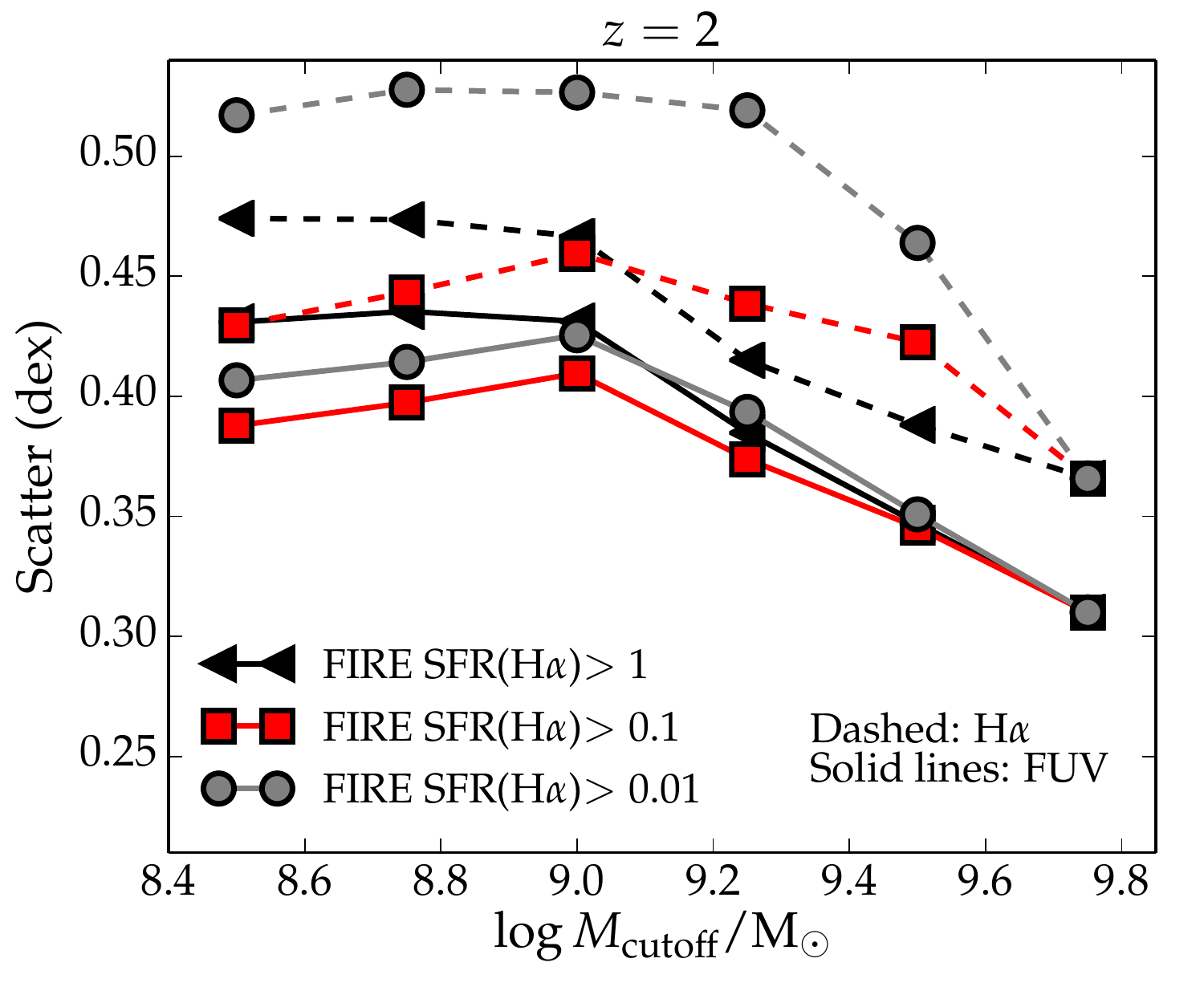}
\caption{
\label{Fig498_SFMSScatter_Mthreshold}
The scatter in the SFR--$M_*$ relation at $z=2$ for galaxies above a stellar mass threshold $M_*>M_\text{cutoff}$. We show the scatter calculated using SFR(H$\alpha$) and SFR(FUV) [\emph{dashed lines} and \emph{solid lines}, respectively]. We show the effect of SFR(H$\alpha$) thresholds of 1, 0.1 and 0.01 $\msun \text{yr}^{-1}$. For all thresholds the H$\alpha$-derived scatter is larger than for FUV, and this is a consequence of bursty star formation. We furthermore see that imposing a H$\alpha$ sensitivity threshold most often makes the scatter decrease.}
\end{figure}

\begin{figure}
\centering
\includegraphics[width = 0.48 \textwidth]{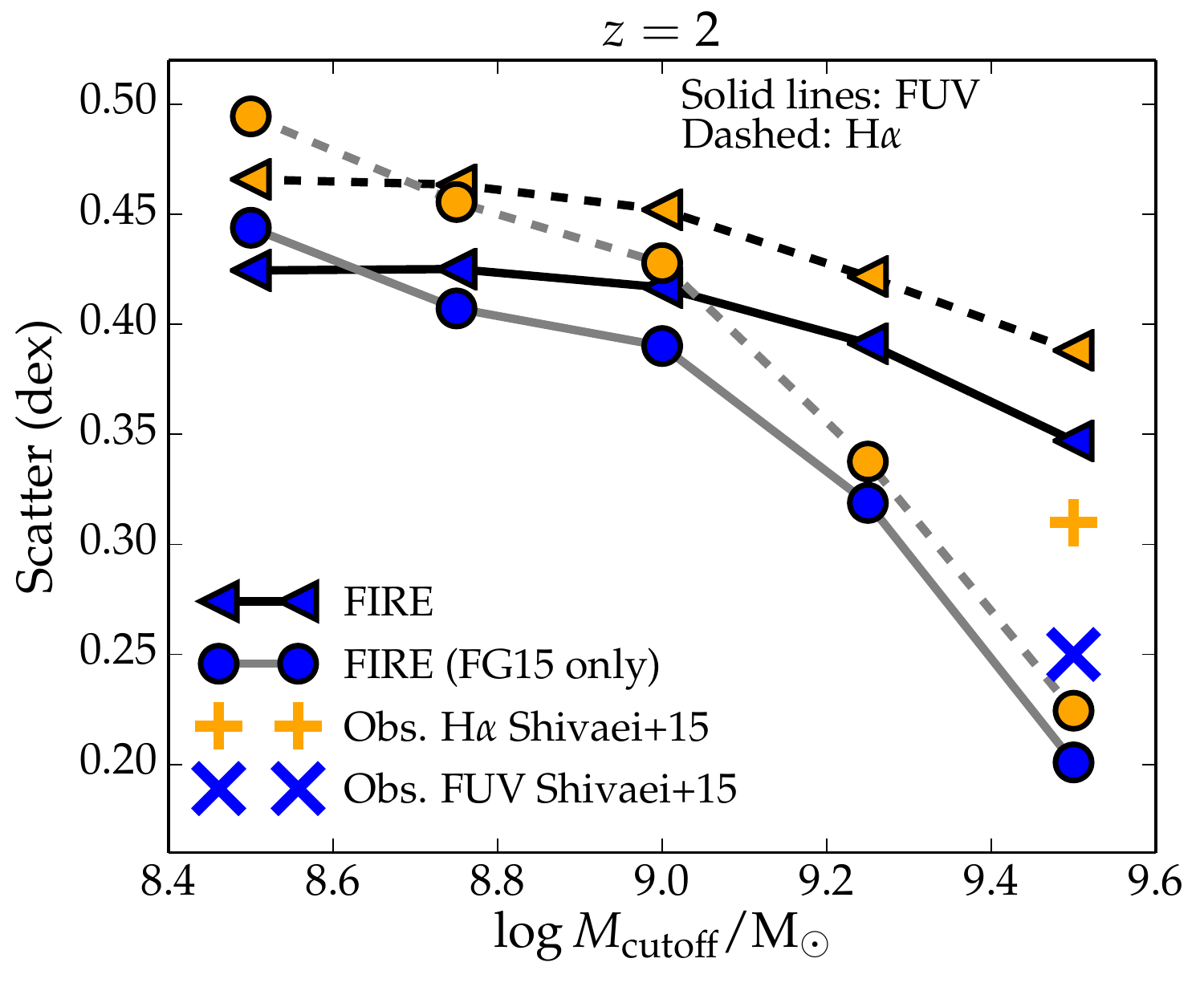}
\caption{
A version of Figure~\ref{Fig498_SFMSScatter_Mthreshold} tailored towards a direct comparison with the observations from \citet{2015arXiv150703017S}. We impose a H$\alpha$-sensitivity of $2 \msun$ yr$^{-1}$ for all lines. We show the scatter for our full sample at $z=2$, and for a sub-sample (FG15) of galaxies in the vicinity of halos with virial masses of $1.9\times 10^{11}-1.2\times 10^{12}\msun$. The difference between H$\alpha$ and FUV is consistent with observations for both our full sample and our sub-sample of galaxies from FG15. The overall level of the scatter in the full sample is larger than observed, and the FG15-sub-sample is more consistent with observations. The difference between the scatter derived between the full sample and the sub-sample of our galaxies hints that systematic effects related to our sample definition are important for the overall value of the scatter.
}
\label{Fig498_SFMSScatter_Mthreshold_SimpleA}
\end{figure}

Real observations of the SFR--$M_*$ relation are only sensitive to a restricted mass range. For example, the $z=2$ observations from \citet{2015arXiv150703017S} only include galaxies with $M_*>10^{9.5}\msun$. To do a realistic estimate of the scatter in this relation, we calculate it for several mass ranges defined by $M_*>M_\text{cutoff}$, where $M_\text{cutoff}$ is the mass cutoff. This is done by first fitting a power law to the SFR--$M_*$ relation, and then the quenched and starburst galaxies are removed from the sample. This removal is done by simply requiring all main sequence galaxies to be within 0.8 dex. The mass-dependence of the scatter is shown in Figure~\ref{Fig498_SFMSScatter_Mthreshold} for both the H$\alpha$- and FUV-derived SFR. We only perform this analysis at $z=2$ because the number of galaxies in our $z=0$ sample is too small to reliably determine the scatter. At $z=2$, we calculate the scatter for SFR(H$\alpha$) limits of 1, 0.1 and 0.01 $\msun \text{yr}^{-1}$ to mimic typical observational H$\alpha$ limits.

The figure reveals that the scatter is larger for the SFR derived with H$\alpha$ than for FUV (independent of the chosen H$\alpha$-sensitivity). This is a direct consequence of bursty star formation histories with rapid SFR-fluctuations on timescales smaller than the sensitivity timescales of the SFR(FUV) indicator. The result that the scatter in the SFR--$M_*$ relation is significantly increased when using the H$\alpha$-based SFR indicator instead of the FUV-based indicator shows that the ISM model in the FIRE simulations is more bursty than the widely used sub-resolution physics models \citep{2003MNRAS.339..289S,2006MNRAS.373.1265O,2012MNRAS.423.1726S,2012MNRAS.426..140D}. Such subgrid models predict very little SFR variability on the short timescales that we are considering here (see Figure~5 of \citealt{2014arXiv1409.0009S}, in which the scatter is almost the same for the instantaneous SFR derived from the gas properties and the 50-Myr-averaged SFR).

It is also visible from Figure~\ref{Fig498_SFMSScatter_Mthreshold} that increasing the H$\alpha$ threshold typically lowers the scatter in the SFR--$M_*$ relation. This is because galaxies that are near the quenching threshold are removed from the sample when the threshold is sufficiently high. An exception to this trend is for the $M_\text{cutoff}\leq 10^{9}\msun$ samples with H$\alpha$ thresholds of 0.1 and 1 $\msun$ yr$^{-1}$, where the former shows the lowest scatter. This situation arises because a threshold of 1 $\msun$ yr$^{-1}$ removes a fraction of the galaxies in the \emph{middle of} the SFR--$M_*$ relation, and the scatter is then calculated for galaxies close to being starbursts.

\subsubsection{A direct comparison to observations and discussion of sample selection effects}\label{EnvironmentalEffects}

In Figure~\ref{Fig498_SFMSScatter_Mthreshold_SimpleA} we directly compare with the observations of \citet{2015arXiv150703017S}, where we impose an SFR(H$\alpha$) threshold of $2 \msun$ yr$^{-1}$. This is identical (within a few per cent) to the $3\sigma$ sensitivity threshold in the $ 2.1 \leq z\leq 2.6$ observations from \citet{2015arXiv150703017S}  (at $z=2.6$, this threshold would be 1.7 times higher than at $z = 2.1$ assuming that the threshold scales with the square of the luminosity distance). We show our full sample of $z=2$ galaxies; this sample exhibits a significantly larger scatter than seen in the observations. We also show a sub-sample of our simulations (from FG15) of halos with virial masses between $1.9\times 10^{11}$ and $1.2\times 10^{12}\msun$, which is at the lower end of our full sample (the full sample is dominated by the MassiveFIRE simulations, which simulate more-massive halos). The scatter for the FG15 sub-sample is significantly lower than for the full sample. This shows that the overall value of the scatter in our simulated sample is significantly affected by sample selection. One effect could be that galaxies in the vicinity of massive halos (such as the MassiveFIRE galaxies) could exhibit greater diversity than galaxies near less-massive halos (such as the FG15 sample). Moreover, the FG15 sample is purely mass-selected, whereas the galaxies in the MassiveFIRE sample are selected to exhibit extreme growth histories. Additionally the density threshold for star formation used in the MassiveFIRE simulations is an order of magnitude lower than in the FG15 simulations. Some or all of these differences may be responsible for the dependence of the scatter on the sample considered. Keeping the above caveats in mind, we note that the scatter observed by \citet{2015arXiv150703017S} is between the value for our full sample and that of the FG15 sub-sample.

Even though the overall level of the scatter is subject to sample selection effects, we still find it meaningful to study the difference in the H$\alpha$- and FUV-derived scatters in our samples. For our full sample, the amount of scatter caused by burstiness is $\sqrt{0.39^2-0.35^2}\simeq 0.17$ dex (for galaxies with $M_*>10^{9.5}\msun$), and for the sub-sample only including the FG15 galaxies, it is $\simeq$0.10 dex. It is hence a robust conclusion that burstiness increases the scatter by around $0.10-0.17$ dex.

This difference in H$\alpha$ and FUV scatter is consistent with the observed difference from \citet{2015arXiv150703017S}\footnote{0.31 dex and 0.25 dex for H$\alpha$ and FUV, respectively.}. The increased scatter in H$\alpha$ compared to UV in these observations could be caused by bursty star formation. However, \citeauthor{2015arXiv150703017S} argues that it is impossible to distinguish whether the difference in the scatter inferred from the two indicators is caused by dust effects, IMF variations and/or observational uncertainties. An example of an observational uncertainty that could play a role is that the FUV flux is measured using imaging, whereas the H$\alpha$ flux is measured using spectroscopy, for which flux calibration is more difficult and slit losses might play a role. When comparing our simulations with observations, we should therefore keep these considerations in mind. The finding that the difference between the H$\alpha$ and FUV scatter predicted by the FIRE simulations is consistent with that observed by \citeauthor{2015arXiv150703017S} is encouraging, but it is uncertain whether the difference in the observed scatter is caused (solely) by bursty star formation histories.

\begin{figure}
\centering
\includegraphics[width = 0.48 \textwidth]{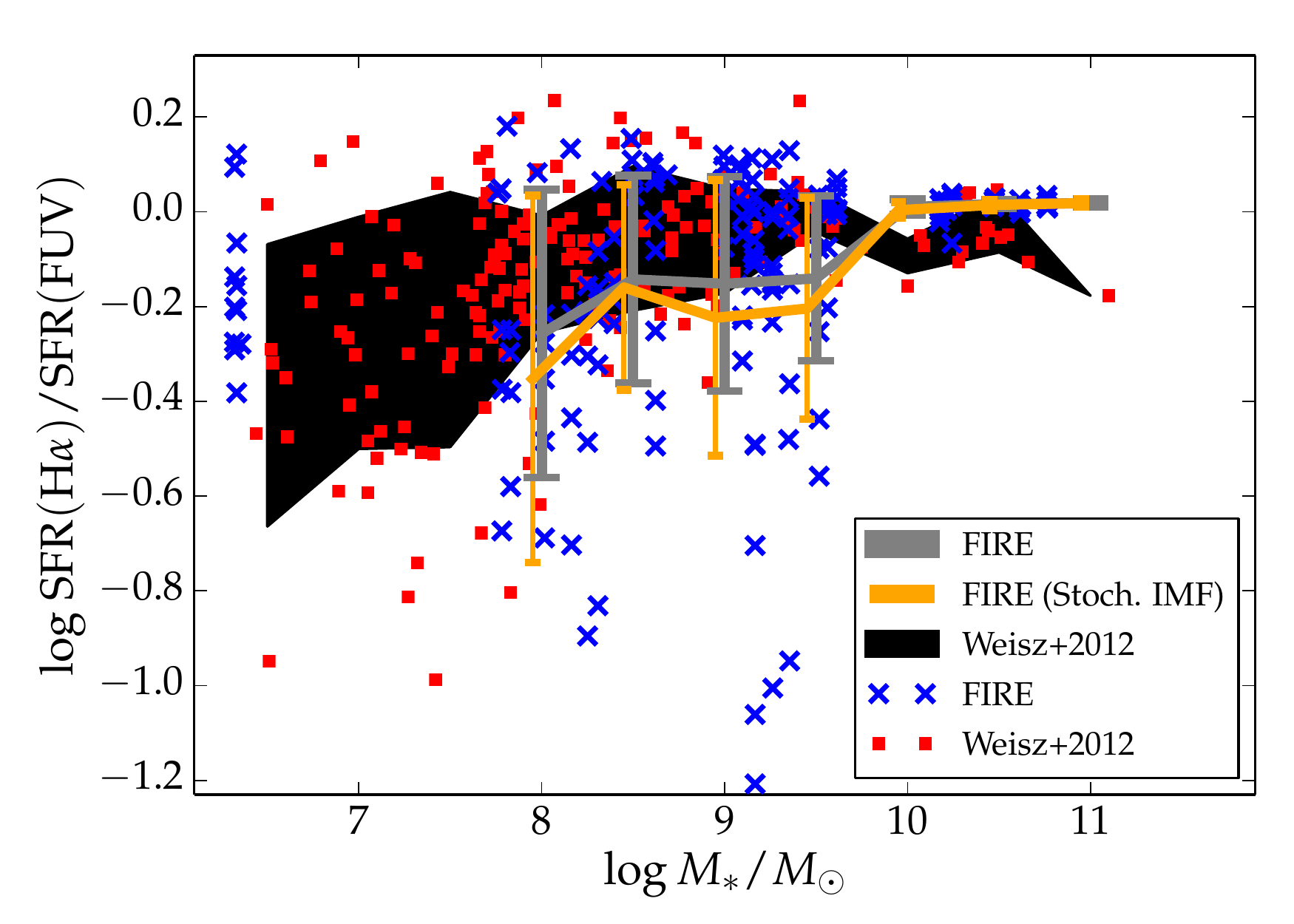}
\caption{Based on the star formation history of the simulated galaxies in FIRE at $z=0, 0.2$, and $0.4$ we have calculated SFR(H$\alpha)/$SFR(FUV) versus stellar mass (\emph{blue $\times$ symbols}). For each galaxy at these redshifts, we show the SFR(H$\alpha$)/SFR(FUV) ratio at lookback times of 0, 25, 50, 75 and 100 Myr. The \emph{grey line} with error bars indicates the median values and 16--84th percentile ranges for different mass bins. We also estimate the role of stochastic IMF sampling in our simulated galaxies (\emph{thin orange error bars} show the 16--84th percentile ranges). Data for local galaxies from \citet{2012ApJ...744...44W} are shown as \emph{red squares}, and the 16--84th percentile ranges for different mass bins are indicated by the \emph{black contour}. The SFR(H$\alpha$)/SFR(FUV) ratios of the galaxies in the FIRE simulations are broadly consistent with the observational data except for the mass range of $10^8 - 10^{9.5} \msun$, but there is a population of simulated galaxies with significantly lower SFR(H$\alpha$)/SFR(FUV) ratios than observed.}
\label{Fig08_Weisz_SLUG}
\end{figure}

\subsection{The distribution of H$\alpha$-to-FUV ratios at $z=0$}\label{ObservationComparison}

We now consider the ratio of the H$\alpha$-derived SFR to the FUV-derived SFR, SFR(H$\alpha$)/SFR(FUV). This ratio is sensitive to the SFR variability of \emph{individual galaxies}, in contrast with e.g. comparisons of the scatter in the SFR--$M_*$ relations obtained using different SFR indicators. Figure~\ref{Fig08_Weisz_SLUG} compares the H$\alpha$ to FUV ratios of 185 local galaxies from \citet{2012ApJ...744...44W} with those of the $z = 0$, $z=0.2$ and $z=0.4$ FIRE galaxies.

For the simulated galaxies, we show the H$\alpha$-to-FUV derived based on the star formation history (\emph{blue} $\times$-symbols and \emph{grey error bars}). We here see that SFR(H$\alpha$)/SFR(FUV) at high masses ($M_*\simeq 10^{10}\msun$) is $\simeq 1$ and the ratio decreases with decreasing stellar mass. Furthermore, the scatter is increased at low stellar masses. This shows how bursty star formation affects the H$\alpha$-to-FUV ratios of galaxies.

\subsubsection{Modeling the effect of stochastic IMF sampling}

The observed H$\alpha$-to-FUV ratios are also affected by other factors than burstiness, such as stochastic sampling of the IMF and dust. To model the effect of stochastic IMF sampling on our simulated galaxies, we re-calculate the H$\alpha$-flux of each simulated galaxy using SLUG including the effects of stochastic IMF sampling. We assume each stellar population particle (with a mass close to the baryonic mass resolution $m_b$) to contain stellar sub-clusters with a stellar mass function of $dN/dM_\text{cl} \propto M_\text{cl}^{-2}$, where $M_\text{cl}$ is the mass of a star cluster. This cluster mass function imposed for each stellar population particle is only non-zero for $20\msun \leq M_\text{cl} \leq m_\text{b}$. The lower limit is the standard value used in SLUG (also used by e.g. \citealt{2011ApJ...741L..26F}). We assume that effects on mass scales greater than the baryonic mass resolution $m_b$ are resolved in our simulations. In Figure~\ref{Fig08_Weisz_SLUG}, we overplot error bars indicating the 16-84\% percentiles of the distribution of the simulated galaxies, where the effect of stochastic IMF sampling is quantified (see \emph{thin orange error bars}). Generally, the scatter in the H$\alpha$-to-FUV ratio is increased by 0.1-0.2 dex, implying that stochastic IMF sampling increases the H$\alpha$-to-FUV ratios of our simulated galaxies. The effect is slightly smaller than that of bursty star formation histories\footnote{At $M_*=10^{8}\msun$, the scatter in the H$\alpha$-to-FUV distribution is increased from 0.3 dex 0.37 dex by stochastic IMF sampling. The amount of scatter induced by this effect is therefore $\sqrt{0.37^2-0.30^2}\simeq 0.22 $ dex, which is slightly smaller than scatter caused by bursty star formation histories.}.

\subsubsection{Is FIRE consistent with observations at $z=0$?}

The scatter at high masses ($M_*>10^{10}\msun$) in the FIRE simulations' SFR(H$\alpha$)/SFR(FUV) ratios is less than the observed scatter, but this may be an artifact of the small number of simulated massive galaxies at $z = 0$. The scatter in the ratio increases with decreasing stellar mass for both the observations and simulations (regardless of whether we include the effects of stochastic IMF sampling). At lower masses, $M_*\lesssim 10^{9.5}\msun$, the scatter is larger than in more-massive galaxies for both the simulations and the observations. A difference between the observations and simulations is that the scatter is slightly larger at low masses in the simulations than in observations.

Overall, the majority of FIRE galaxies have SFR(H$\alpha$)/SFR(FUV) ratios consistent with observations, but a fraction of the simulated galaxies do have significantly lower ratios than observed (even when we do not include the effects of stochastic IMF sampling). These are galaxies in strong post-burst epochs. This makes the scatter in SFR(H$\alpha$)/SFR(FUV) of the simulated galaxies around 0.2 dex larger than the observations for $M_*\lesssim 10^{9.5}\msun$. When we include the effect of stochastic IMF sampling, this difference in scatter is increased to around 0.3 dex. We conclude that a fraction of our $z = 0$ simulated galaxies with $M_*\lesssim 10^{9.5}\msun$ have slightly lower SSFRs (or equivalently H$\alpha$ equivalent widths) than real galaxies, even though the majority of our galaxies are consistent with observations.

The H$\alpha$-to-FUV ratios can also be used to characterise the burst cycles in the simulations at $z=2$, but the lack of observations with very deep SFR(H$\alpha$) limits makes a direct comparison with observations impossible. We thus discuss the H$\alpha$-to-FUV ratios of $z=2$ galaxies in Appendix~\ref{HalphaFUVRedshift2}.

As noted in \citet{2015arXiv150703017S}, the method used to correct for dust might also affect the H$\alpha$-to-FUV ratios. Currently, there is no consensus about whether the observed scatter in H$\alpha$/FUV is caused by bursty star formation, dust effects or /andstochastic IMF sampling, so it is unfortunately not possible to perform a completely robust comparison between simulations and observations. A more complete analysis, including performing radiative transfer on the simulated galaxies to directly calculate observed SFR indicators rather than dust-free ones \citep[e.g.][]{2014arXiv1402.0006H}, would yield a more direct comparison between our simulations and observations.

\begin{figure*}
\centering
\includegraphics[width = 0.83 \textwidth]{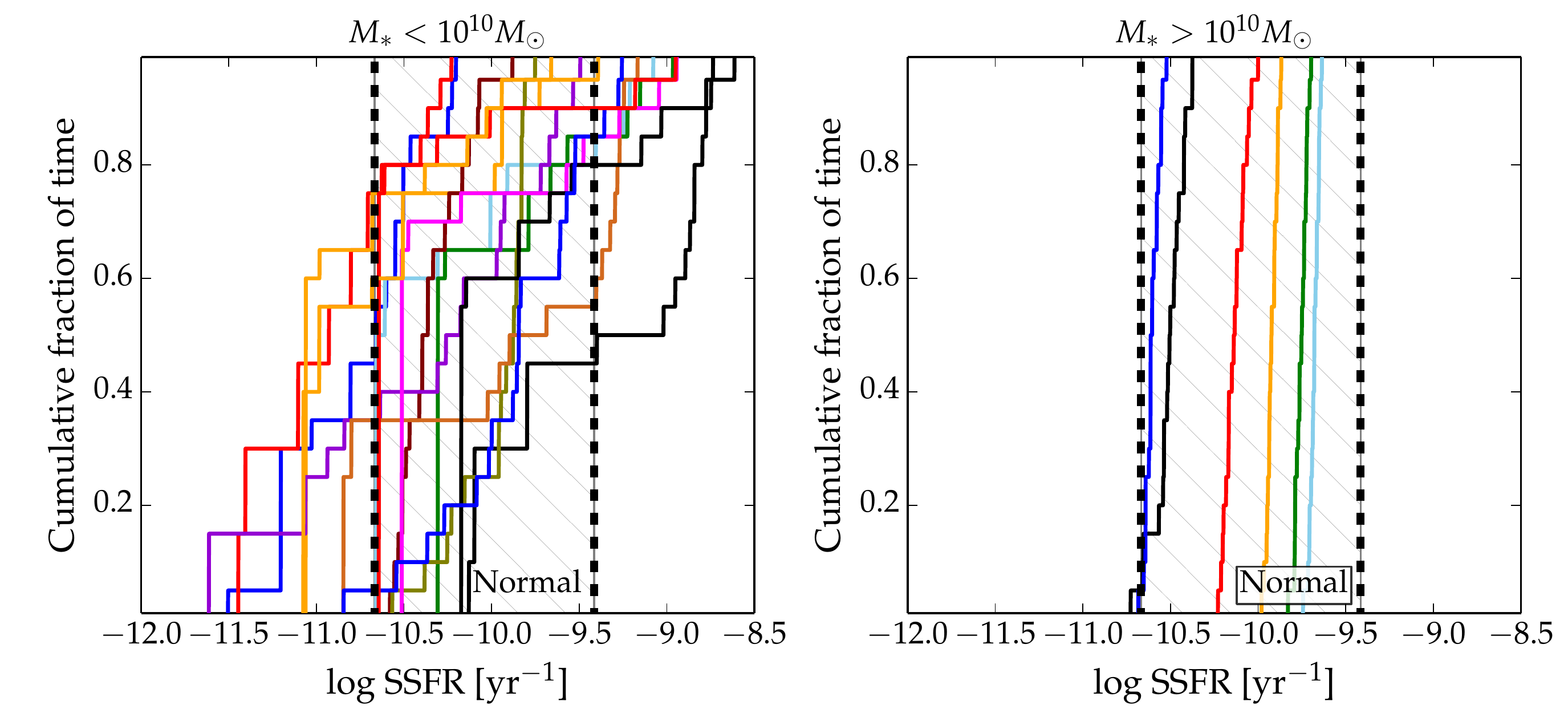}
\caption{Cumulative distribution functions of the 10-Myr-averaged specific SFR, S\sfrten $\equiv \text{SFR}_\text{10 Myr}/M_*$, for a subset of the galaxies at $z=0,0.2$ and 0.4 in our simulations (we do not mark whether a line corresponds to a galaxy analysed at $z=0$, $0.2$ or $0.4$).
The S\sfrten values were calculated for 20 10-Myr non-overlapping intervals within the last 200 Myr of a galaxy's history.
The \emph{dashed vertical lines} enclose a SSFR-interval, where a galaxy is classified as normal (or `main sequence') according to a SFR-indicator with a sensitivity timescale of 200 Myr. The \emph{left panel} shows galaxies with
$M_*<10^{10}\msun$, and the \emph{right panel} shows galaxies with $M_*>10^{10}\msun$. In the \emph{left panel}, 15 randomly selected galaxies
are shown. At $z \simeq 0$, low-mass galaxies have larger SFR fluctuations than
more-massive galaxies. In terms of \sfrten, \emph{a low-mass galaxy's SSFR can vary by more than two orders of magnitude within 200 Myr}.
The galaxies with $M_* > 10^{10} \msun$ are exclusively `main sequence' galaxies.}
\label{Fig06B_SSFRHist}
\centering
\includegraphics[width = 0.83 \textwidth]{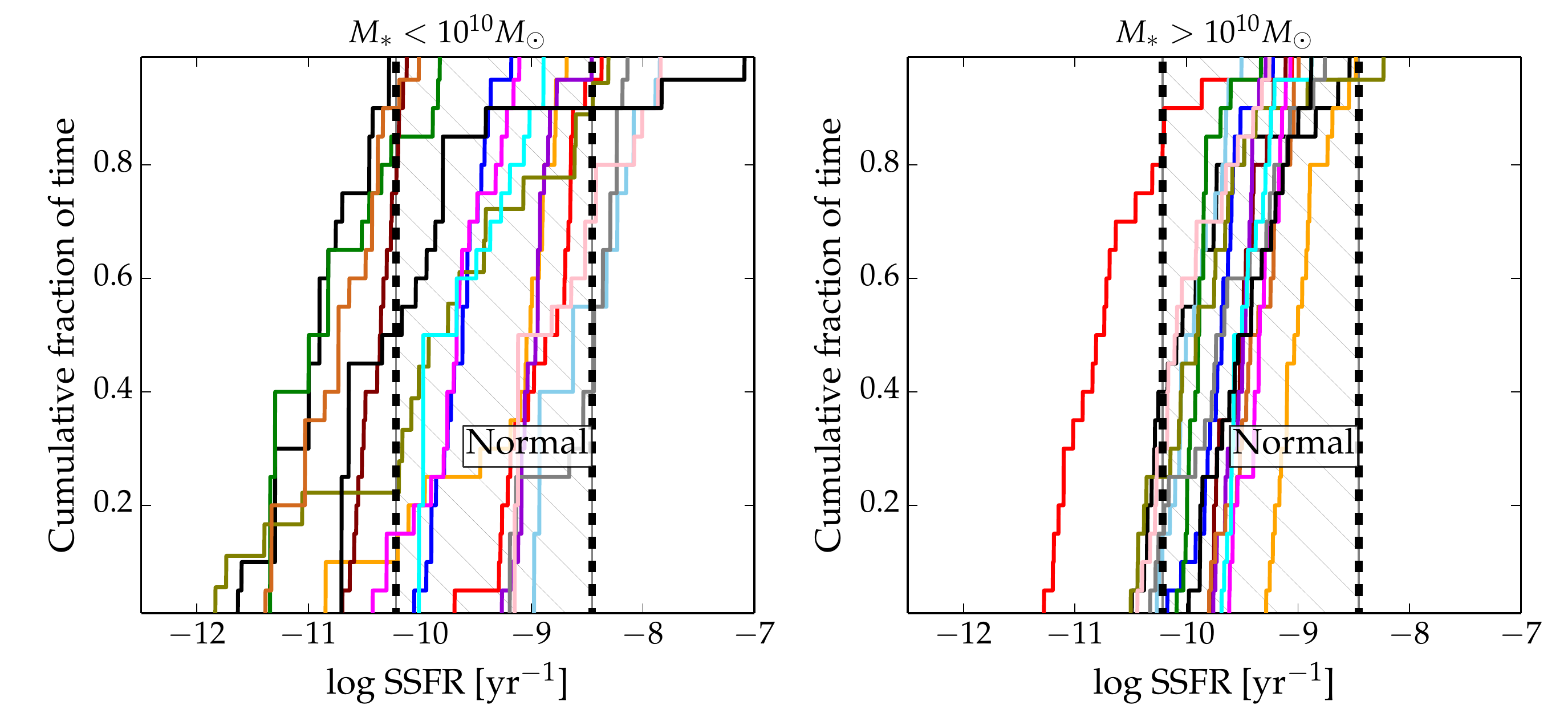}
\caption{Similar to Figure~\ref{Fig06B_SSFRHist}, but for $z=2$. Only 15 galaxies are shown in each panel. At this redshift, the high-mass galaxies' SSFRs
vary more than at $z = 0$, but the low-mass galaxies still exhibit greater variations in \sfrten over 200-Myr intervals.}
\label{Fig06B_SSFRHist_z2}
\end{figure*}

\begin{figure*}
\centering
\includegraphics[width = 0.95 \textwidth]{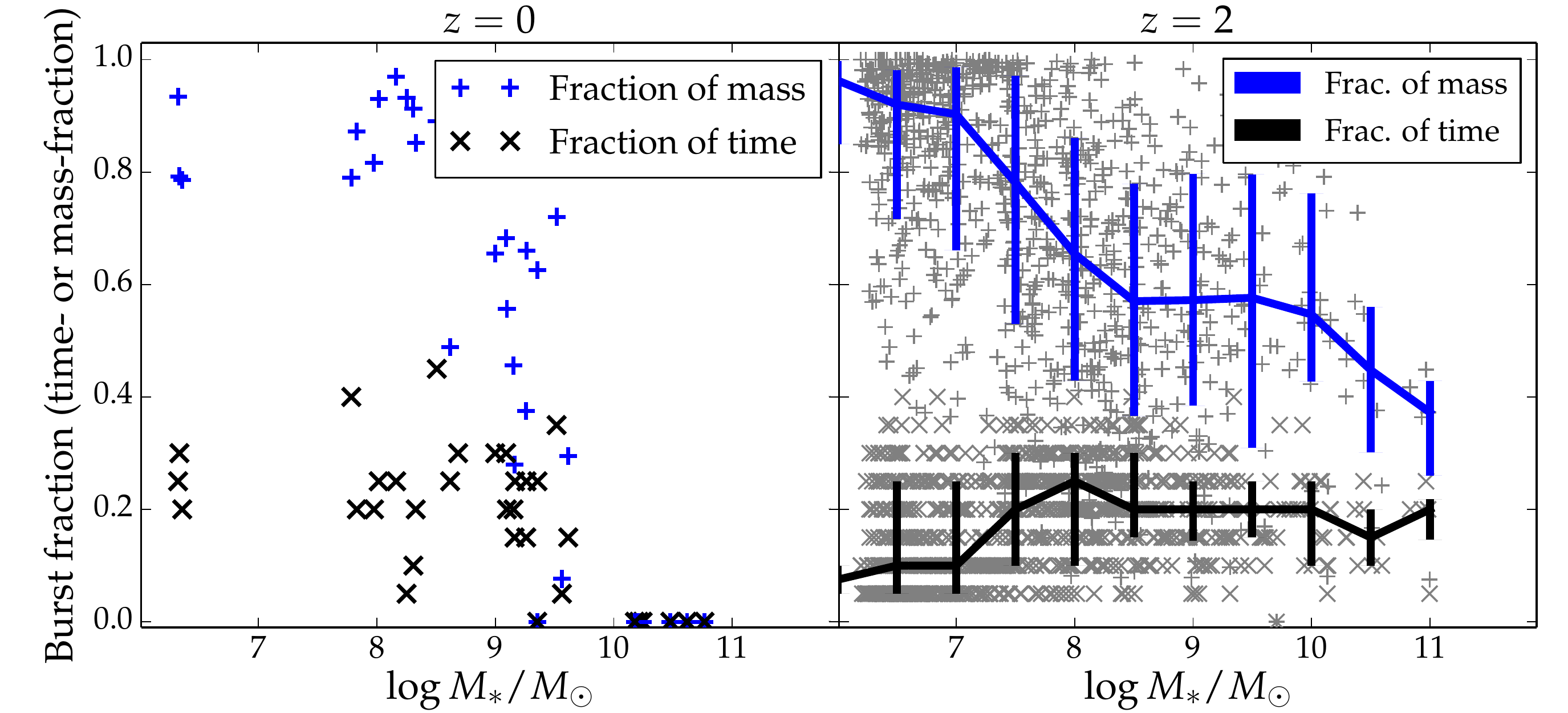}
\caption{The fraction of time spent ($\times$ symbols) and the fraction of stellar mass formed ($+$ symbols) in burst cycles. Again, a galaxy is defined to be in a burst phase when SFR$_\text{10 Myr}>1.5\times$SFR$_\text{200 Myr}$. The \emph{left} panel shows galaxies at $z=0,0.2$ and 0.4, and the \emph{right} panel shows $z=2$ galaxies. For the $z=2$ galaxies, the 16--84th percentile ranges are indicated by the error bars. At $z = 0$, galaxies with $M_*\lesssim10^{9}\msun$ spend $\simeq 25$\% of their time and form $\simeq 80$\% of their stars in burst cycles. More-massive galaxies are not bursty. At $z = 2$, the fraction of time spent in bursts (the \emph{duty cycle}) is $\simeq 20$\%, independent of the galaxy mass. The fraction of the stellar mass formed in bursts decreases from $\simeq 100$\% at the lowest masses to $\simeq 40$\% at $M_* = 10^{11} \msun$.}
\label{Fig06A_BurstSFRFrac}
\end{figure*}

\subsection{Galaxies going through burst cycles}\label{SFMSBurst}

Having studied the SFR-$M_*$ relation and its scatter, we will now study in more detail the presence of short ($\sim$10 Myr, corresponding to the timescale traced by H$\alpha$ emission) SFR fluctuations within a 200-Myr time interval (i.e. the approximate timescale traced by FUV emission). In Figure~\ref{Fig06B_SSFRHist}, we show how SSFR$\equiv$SFR$_\text{10 Myr}/M_*$ of individual galaxies varies over a 200-Myr time interval at $z=0$. Each curve shows the cumulative distribution of the \sfrten values of a single galaxy calculated in 20 non-overlapping 10-Myr time intervals. The galaxies with $M_*>10^{10}\msun$ (\emph{right panel}) exhibit a relatively small amount of SFR variability, and individual galaxies' SFR$_\text{10 Myr}$ values do not change by more than a factor of three over the 200-Myr interval.

As a comparison, the \emph{dashed vertical lines} in Figure \ref{Fig06B_SSFRHist} indicate the width of the main sequence based on the 200-Myr SFR indicator. The width is selected to be $2.5 \sigma$, with $\sigma = 0.3$ dex; this is an observationally motivated choice. (Recall that we have already demonstrated that the main sequence scatter determined using a 10-Myr SFR indicator is greater than for a 200-Myr indicator.) We see that for the simulated galaxies with $M_*>10^{10}\msun$, the 10-Myr-averaged SFR is essentially always characterised as normal according to the main sequence scatter on a 200-Myr timescale because for these galaxies, there is little SFR variability on 10-Myr timescales. In contrast, for low-mass galaxies ($M_*<10^{10}\msun$), the 10-Myr SFR can vary from below to above the `main sequence' (defined based on the 200-Myr SFR indicator)
within a 200-Myr period.

Figure~\ref{Fig06B_SSFRHist_z2} shows the same plots but for $z = 2$. For visibility reasons (including all our $>900$ galaxies at $z=2$ would make the plot unreadable), we have down-sampled\footnote{This down-sampling is done by selecting random numbers to decide which galaxies to include. The trends found are robust to the sampling.} the number of galaxies, so we only show 15 galaxies in each panel. As at $z = 0$, high-mass galaxies exhibit less SFR$_\text{10 Myr}$ variability than low-mass galaxies. However, $M_* > 10^{10} \msun$ galaxies exhibit more \sfrten variability at $z = 2$ than at $z = 0$.

An important thing to keep in mind is that different SFR indicators have different quenching and starburst thresholds because the scatter of the SFR--$M_*$ relation depends on the SFR indicator. When \sfrten is less than the quenching threshold for the 200-Myr SFR indicator, a galaxy is thus not necessarily permanently quenched; rather, it is more likely to only have a temporarily low \sfrten-value. At $z = 2$, even massive ($M_* > 10^{10} \msun$) galaxies can have \sfrten values classifying them as both quenched and `main sequence' within a 200-Myr time interval according to the quenching thresholds derived for the \sfrth indicator. Thus, at $z \simeq 2$, galaxies may be mistaken as quenching or quenched galaxies when in reality, they will be forming stars again with a high \sfrten within 100 Myr. \citet{2015arXiv150307164B} presented observations of GDN-8231, a $M_*=6\times 10^{10}\msun$ galaxy at $z=1.7$ with a young stellar population with an age of 750 Myr and H$\alpha$- and 24-\micron-derived SFRs of $\lesssim 10 \msun \text{yr}^{-1}$. They interpreted these observations as evidence that the galaxy is `caught in the act' of quenching. However, both H$\alpha$- and 24-\micron ~emission are short-timescale SFR indicators. Our simulations show that in terms of S\sfrten, even massive galaxies at $z = 2$ may be classified as both quenched and normal star-forming galaxies within 200 Myr. Thus, perhaps GDN-8231 has been observed in a temporary phase of low S\sfrten and is not in fact permanently quenched.

\subsection{Dividing a star formation history into post-burst, steady and burst phases} \label{BurstySFH}

The terms post-burst and burst refer to a galaxy's current star-formation rate being lower and higher, respectively, than some measure of its SFR in the recent past. We now choose a definition of burstiness that quantifies how actively star-forming a galaxy has been in the last 10 Myr of its lifetime compared to the last 200 Myr. Specifically, we define a galaxy to be in burst, post-burst or steady phases based on the following criteria:

\begin{align*}
&\text{Burst phase: } \text{SFR}_\text{10 Myr}>1.5 \times \text{SFR}_\text{200 Myr}, \\
&\text{Post-burst phase: } \text{SFR}_\text{10 Myr}<\frac{\text{SFR}_\text{200 Myr}}{1.5}, \\
&\text{Steady phase: } \frac{\text{SFR}_\text{200 Myr}}{1.5}<\text{SFR}_\text{10 Myr}<1.5 \times \text{SFR}_\text{200 Myr}.
\end{align*}

The factors of 1.5 in these definitions are arbitrary. In observational applications, one approach would be to study the statistical behaviour of SFR(H$\alpha$)$/$SFR(FUV), which characterises whether a galaxy is undergoing a burst (see Section \ref{SFRIndicatorIntro}) for a large sample of galaxies and consequently infer the roles of the burst, post-burst and steady phases of star formation histories. Another approach is to use a combination of the 4000-\AA ~break and the Balmer absorption-line index H$\delta_A$ to constrain the mass fraction formed in a recent burst \citep{Kauffmann2003}. However, the focus of the current analysis is to theoretically illustrate how stars are formed in simulations, which is clearer if we use the above definitions. In Section \ref{ObservationComparison}, we will provide a direct comparison with observations in terms of the SFR(H$\alpha$)$/$SFR(FUV) ratio.

Given the large amplitudes and frequency of the starbursts experienced by the FIRE galaxies, it is worth considering how much stellar mass is formed in bursts. Figure~\ref{Fig06A_BurstSFRFrac} shows the fraction of stars formed (blue crosses) and time spent (black $\times$'s) in burst phases for $z = 0$ (\emph{left panel}) and $z = 2$ (\emph{right panel}).  At $z = 0$, low-mass galaxies ($M_*<10^{9} \msun$) form most of their stars ($\ga80$\%) in bursts, but they spend a relatively small fraction of their time (15-35\%) in burst phases. As expected from the above discussion, both fractions are zero for galaxies with $M_* > 10^{10} \msun$. At $z = 2$, the lowest-mass galaxies form effectively all of their stars in bursts despite spending $\la 20$\% of their time in bursts; the reason is that most of their time is spent in post-burst phases, in which their SFRs are extremely low. Although massive galaxies spend a similar (or even slightly greater) fraction of the time in bursts, the fractions of their stars formed in bursts are less than for low-mass galaxies ($\la 50$\% for the most-massive galaxies) because the massive galaxies spend a large fraction of their time in steady phases, in which the SFRs are less than in the burst phases but still sufficiently high to account for a significant fraction of the stellar mass formed over the 200-Myr interval. Still, the fact that massive galaxies at $z = 2$ form approximately half of their stars in bursts is in marked contrast with $z = 0$, where massive galaxies form effectively no stars in bursts.

The behaviour of galaxies with $M_*> 10^{10}\msun$ from the MassiveFIRE simulations suite was also studied in \citet{2016arXiv161002411F}, who also noted the presence of short bursts of star formation. Additionally, galaxies often went through a temporarily suppressed star formation state immediately after the burst.

\begin{figure*}
\centering
\includegraphics[width = 0.97 \textwidth]{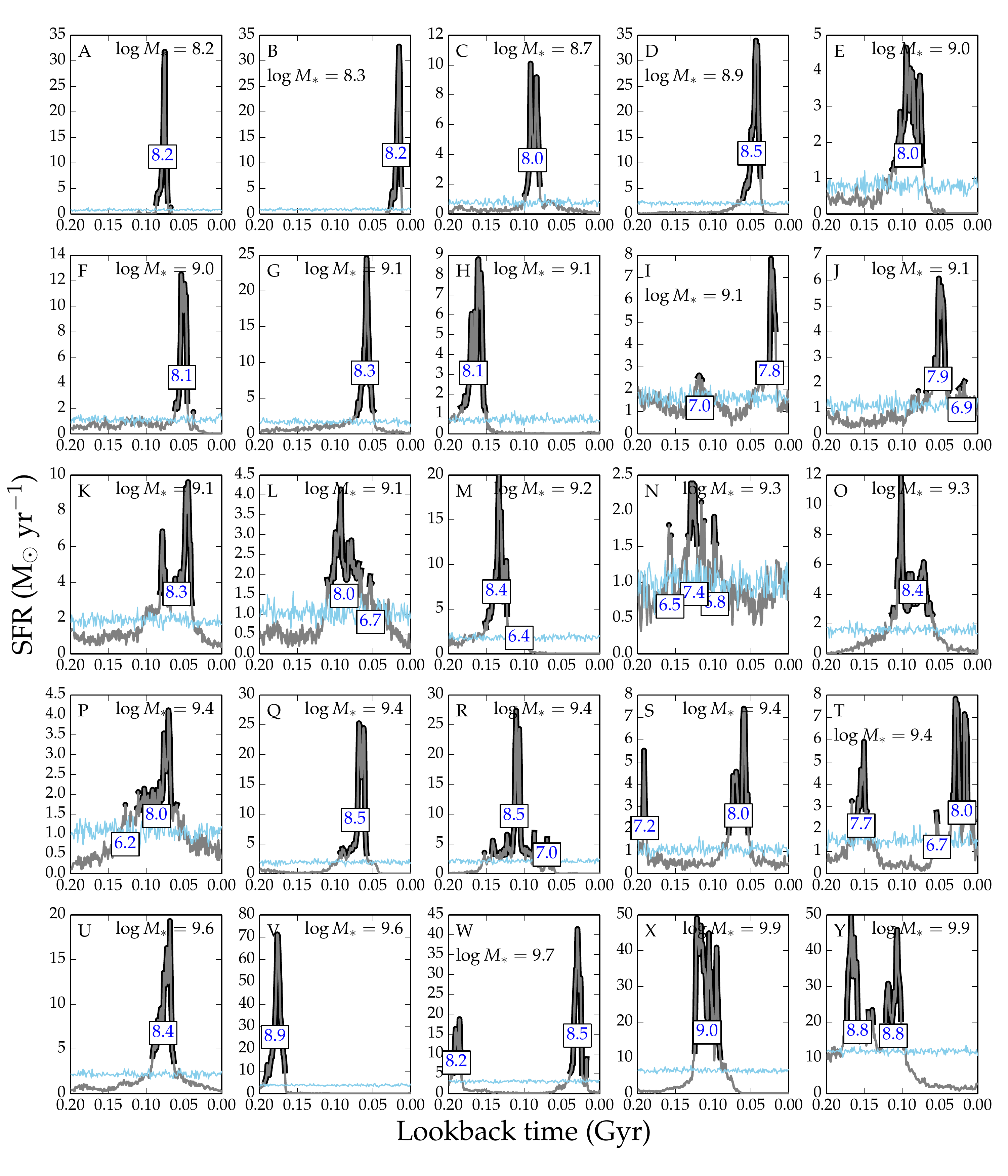}
\caption{Each panel shows the last 200 Myr of the star formation history of a galaxy at $z=2$ (\emph{grey line}) binned in 1-Myr time bins. A galaxy is defined to be in a burst phase (marked by \emph{thick black lines}) when the SFR is greater than 1.5 times the average SFR over the last 200 Myr. The \emph{blue numbers} in the boxes show the log of the stellar mass formed within one burst period consisting of consecutive points obeying this criterion. Some burst periods form up to around $10^9 \msun$ of stars. An example of how the SFH would look if all of the variability were due to Poisson noise of the sampling of star particles is shown by the \emph{thin blue line}. The stellar masses in solar units (at the end of the time intervals) are shown in the \emph{upper right corners}. The figure shows that the shortest bursts in the FIRE simulations have durations of 5-50 Myr, and the SFR peaks typically have durations as short as 3 Myr.}
\label{Fig321_MassiveFIRE_1Myr}
\end{figure*}

\begin{figure*}
\centering
\includegraphics[width = 0.97 \textwidth]{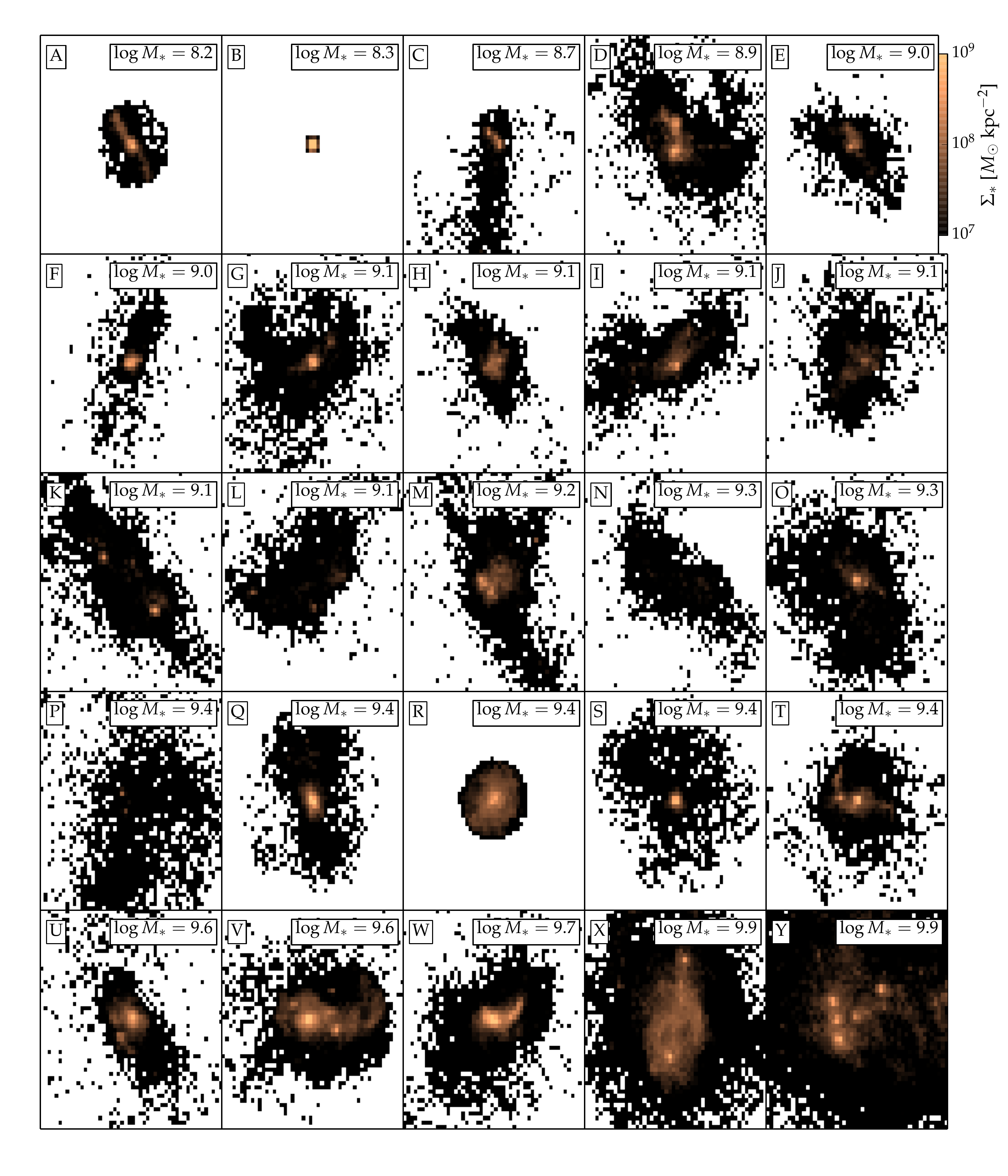}
\caption{The surface density distribution of the stars formed in bursts according to the burst definition illustrated in Figure~\ref{Fig321_MassiveFIRE_1Myr}. Each panel has dimensions of 10 kpc$\times$10 kpc. Many galaxies have a single strong peak in surface density (panels A, B, C, E, F, G, H, I, O, Q, R, and S), but others have their stars formed in multiple regions (most clearly in panels D, M, U, V, W, X, and Y). The five most-massive galaxies have more star-forming clumps than the five least-massive galaxies. The same logarithmic colour coding is used in all panels (see \emph{colour bar} in the \emph{upper right panel}).}
\label{Fig321_MassiveFIREA_1MyrSpatial}
\end{figure*}

\subsection{The variability timescale and duration of burst cycles}\label{BurstsTimescale}

Until now, we have studied the SFR variability on timescales equal to or longer than 10 Myr. Despite being observationally inaccessible, variability on shorter timescales might play an important role in shaping galaxies. In Figure~\ref{Fig321_MassiveFIRE_1Myr}, we plot the SFR in bins of width 1 Myr for 25 randomly chosen galaxies with $M_*<10^{10}\msun$ from the MassiveFIRE suite.
To obtain well-sampled star formation histories, we select galaxies that have formed more than 5000 star particles over the last 200 Myr within 20\% of the virial radius of the halo. This corresponds to the requirement that \sfrth $\ga 1 \msun$ yr$^{-1}$. Given that our plotted galaxies have $10^8\msun \lesssim M_* \lesssim 10^{10}\msun$, this SFR threshold corresponds to starburst galaxies at $M_*=10^8 \msun$ and normal galaxies for $M_*\simeq 10^{10}\msun$ (see the plot of the SFR--$M_*$ relation in Figure~\ref{Fig01_SFMS}). Consequently, especially for low-mass galaxies, the bursts will be stronger than for normal galaxies at that mass. We mark galaxies as being in a burst phase (\emph{thick black lines}) when the SFR is at least 1.5 times the \sfrth value at $z=2$.
Some of the shortest bursts are shown in panels A, B, D, F and V where the SFR exhibits a single peak, and before (after) the peak the SFR increases (decreases) monotonically.
In all of these cases, the burst peak is resolved by at least three time bins, which implies that the shortest variability timescales of bursts are of order 3 Myr. The most common type of
bursts have longer durations and more complex shapes; see e.g. panels E, H, K, L, X and Y. The typical burst durations in these panels are 25-50 Myr, but some short spikes have durations as short as 3-5 Myr. 

The presence of SFR variability on timescales as small as 3 Myr suggests that the FIRE feedback model leads to SFR fluctuations that cannot be probed using standard SFR indicators such as H$\alpha$ and FUV emission.\footnote{In principle, these fluctuations could be probed for local galaxies by analysing their resolved stellar populations
\cite[for recent examples of such analyses, see e.g.][]{Weisz08,Weisz11,Weisz14,Johnson13,Williams15}.}
An important consequence of such fast SFR variability is related to the inner density profiles of dark matter halos because SFR variability on timescales less than the local orbital period of dark matter particles can turn dark matter cusps into cores (\citealt{2012MNRAS.421.3464P}, see also \citealt{2013MNRAS.433.3539G,2014MNRAS.437..415D,2015arXiv150202036O,Chan2015,2015arXiv150804143R}).

\begin{figure*}
\centering
\includegraphics[width = 0.95 \textwidth]{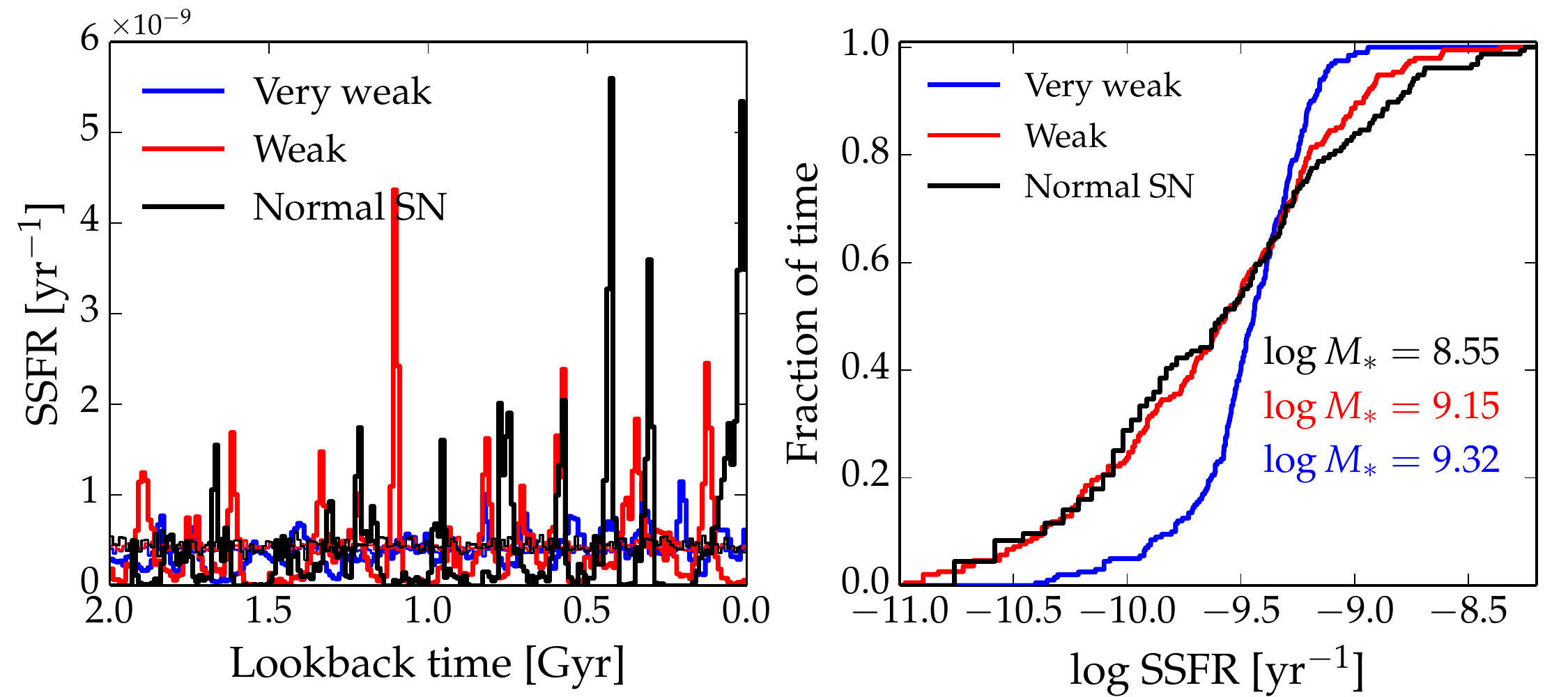}
\caption{This figure shows how decreasing the role of supernova type II feedback changes the burstiness of the star formation history of the m12v halo's central galaxy at $z=2$. We here define SSFR$\,\equiv\,$SFR$_\text{10 Myr}/M_*(z=2)$. \emph{Left panel}: The star formation histories for the m12v galaxy simulated with normal, weak and very weak Type II supernova feedback. Making supernova feedback weaker results in less bursty star formation histories. This confirms our intuition that violent supernova feedback is one of the main causes of the burstiness of the FIRE galaxies' star formation histories.
}
\label{Fig714_PhysVar_SFH}
\end{figure*}

In Figure~\ref{Fig321_MassiveFIRE_1Myr}, the amount of stellar mass formed in burst epochs is written with \emph{blue} numbers in boxes. The stellar mass formed in such burst epochs spans a mass range of $10^{6.2}-10^{9.0}\msun$. In Figure~\ref{Fig321_MassiveFIREA_1MyrSpatial} we plot the surface density of stars formed during the bursts. We perform a projection into a 10 kpc$\times$10 kpc plane and calculate the surface density on a grid with 50$\times$50 bins. The most-massive galaxies ($M_*> 10^{9.5}\msun$) form their stars in several regions; panel Y, for example, reveals five star-forming clumps. At lower stellar masses (panels A-J), the bursts of star formation typically occurs in single clumps. In contrast, star formation in massive galaxies is distributed over a larger area in several star-forming clumps. This is consistent with the result that star formation becomes less bursty with increasing stellar mass.

We conclude that our selection of galaxies typically form their stars in clumps of $10^{6.2}-10^{9.0}\msun$. The upper limit is very conservative, because bursts occur in several clumps for the most massive galaxies; thus, the maximum mass of a clump is probably $2-5$ times smaller than $10^9\msun$. These clump masses are larger than for those of GMCs in the Milky Way, which usually form stellar masses of up to a few times $10^5 \msun$ \citep{2011ApJ...729..133M}. However, recall that our analysis is performed at $z=2$, and $z \sim 2$ galaxies exhibit star-forming clumps that are much more massive than in the Milky Way \citep[e.g.][]{2011ApJ...739...45F}. \citet{2016arXiv160303778O} demonstrated that the clump properties (e.g. masses) of our simulated galaxies are consistent with those of $z \sim 2$ clumpy discs. Also, the low-mass galaxies shown here are not typical star-forming galaxies but rather extreme starbursts (because of our requirement that at least 5000 stellar population particles are formed in the last 200 Myr before $z=2$), in which a nuclear concentration of intense star formation is expected.

\section{Supernova feedback as a driver of burstiness} \label{SNSuite} \label{Modboost}

To explicitly demonstrate that Type II supernova feedback plays a dominant role in determining the burstiness of the FIRE galaxies' star formation histories, we have performed two additional simulations of the m12v halo with supernova feedback that is weaker than in our fiducial simulation. We will examine three different simulations of the m12v halo at $z=2$ from \citet{2013arXiv1311.2073H}. We will use runs with normal supernova feedback (\emph{Normal SN}), \emph{weak feedback} and \emph{very weak feedback}. For the \emph{weak} and \emph{very weak} runs, the SN feedback is artificially weakened by decreasing the momentum deposited into the surrounding gas by factors of 4 and 8, respectively.

The \emph{left panel} of Figure~\ref{Fig714_PhysVar_SFH} shows SSFR$\,\equiv\,$SFR$_\text{10 Myr}/M_*(z=2)$ versus time for the main m12v halo in the three different simulations, and the \emph{right panel} shows the cumulative distribution functions of the SSFR for each simulation. Because the galaxies in the FIRE simulations evolve stochastically, it is only meaningful to compare the simulations in a statistical manner (i.e. comparing the SSFR values at a fixed time is not useful). These plots reveal that when the supernova feedback is stronger, the variations in the SSFR are greater: not only is the SSFR less in the post-burst phases, but also the SSFR is greater during the bursts. Thus, this test clearly indicates that supernova feedback is one of the primary causes of the burstiness of the FIRE galaxies' star formation histories, and it can result in bursty SFHs even in massive galaxies at $z \sim 2$. However, the simulated massive ($\mstar \gtrsim 10^{10} \msun$) galaxies at $z \lesssim 1$ exhibit relatively smooth SFHs, perhaps because supernova feedback is unable to drive strong outflows (and subsequently galactic fountains) in such galaxies \citep{2015arXiv150103155M,HH2015}.

It is intuitive that stronger SN feedback results in lower SSFRs in post-burst phases, but the fact that the maximum SSFR is increased may seem counter-intuitive. There are (at least) two possible reasons for this effect: (1) the stronger SN feedback causes more gas to be kicked out of the galaxy but not the halo, resulting in more prominent galactic fountains. This gas rains down on the galaxy at a later time, resulting in a higher SSFR than would occur if the SN feedback were weaker. (2) When the SN blastwaves interact with the ambient ISM, the resulting shock compression could cause triggered star formation. A detailed analysis of these two possibilities is beyond the scope of this paper.

\section{Discussion} \label{Discussion}

\subsection{What we can learn from the H$\alpha$-to-UV ratio}

Understanding the observed distribution of the H$\alpha$-to-UV ratio is a challenging problem because several physical mechanisms might affect this ratio. Of importance are naturally the mechanisms that alter the fraction of short-lived massive stars, which include stochastic IMF sampling, IMF variations and bursty star formation histories \citep{2009ApJ...706..599L,2011ApJ...741L..26F,2012MNRAS.422..794E,2012ApJ...744...44W, 2014MNRAS.444.3275D}. Additionally, dust attenuation also influences the observed ratio because the UV flux is attenuated more than the H$\alpha$ flux \citep[see the discussion in][]{2009ApJ...706..599L}. Disentangling the roles of each of these effects is very difficult, but the observed ratio can still provide important constraints on each process, as we saw in Section~\ref{ObservationComparison}.

Because one of the main drivers of burstiness in the FIRE simulations is supernova feedback (Figure~\ref{Fig714_PhysVar_SFH}), alternate implementations of supernova feedback could affect the resulting H$\alpha$-to-UV ratios of simulated galaxies. There is, however, not much room for modifying the supernova feedback coupling in the FIRE physics model. In dwarf galaxies, individual supernova remnants are resolved, and even in the more massive galaxies, they are time-resolved, and the subgrid model should do a reasonable job at capturing the unresolved phases \citep[the uncertainties are at the tens of percent level; ][]{2015MNRAS.450..504M}. Another argument against supernova feedback being the relevant mechanism to modify is that decreasing the strength of this feedback (from the \emph{Normal SN} to the \emph{Weak} run in Figure~\ref{Fig714_PhysVar_SFH}) had little influence on the SFR variability.

An additional physical effect that could be implemented in simulations is stochastic IMF sampling (see \citealt{Cervino2013} for a review). We have shown that accounting for this effect in our stellar population synthesis calculations decreases the mean and median H$\alpha$-to-UV ratios and also alters the scatter in the ratio \citep[see also ][]{2014MNRAS.444.3275D,2016arXiv160405314G}. Stochastic sampling of the IMF could also alter the effectiveness of stellar feedback. The momentum and energy imparted by radiative feedback from massive stars is implemented assuming full IMF sampling and thus can be `diluted' on average in low-mass galaxies. By this, we mean that when a star particle is spawned, if it is not sufficiently massive, the momentum and energy deposited in an IMF-averaged way in our simulations can sometimes be less than those associated with a single massive star which could in reality form. If stochastic IMF sampling were implemented, the energy and momentum deposition could occur in a more bursty manner; the effects of stochastic IMF sampling on feedback will be discussed in detail in a forthcoming work (K.-Y. Su et al., in prep.)

The observed H$\alpha$/FUV ratio is also sensitive to the stellar population synthesis models employed. For example, inclusion of binaries can extend the lifetimes of massive stars that produce ionizing photons \citep{2016arXiv160107559M} and potentially lead to reduced scatter in the H$\alpha$/FUV ratios predicted for the simulations. Another effect that could cause increased burstiness in the simulations is that we do not fully resolve the GMC mass function in our simulations. It is possible that not resolving low-mass GMCs could cause of at least some of the apparent discrepancies between the simulations and observations, including the relatively large fraction of temporarily quenched galaxies and the $z \simeq 0$ simulated galaxies with lower SFR(H$\alpha$)/SFR(FUV) ratios than observed (Figure \ref{Fig08_Weisz_SLUG}). If the GMC mass function were fully resolved, the lowest-mass GMCs, which dominate in terms of number but not mass, might provide a relatively constant minimum SFR. Consequently, higher-resolution simulations may exhibit minimum SFR(H$\alpha$)/SFR(FUV) values greater than those of the present simulations. This would both decrease the fraction of temporarily quenched galaxies and decrease the scatter in the SFR(H$\alpha$)/SFR(FUV).

A limitation of using the H$\alpha$-to-FUV flux ratio to constrain bursty star formation is that this ratio is difficult to constrain accurately for individual galaxies even at $z=0$, and it is much harder to measure for high-redshift galaxies. Luckily, new instruments, such as the Multi-Object Spectrometer for Infra-Red Exploration \citep[MOSFIRE;][]{McLean2012} at the Keck Observatory, are making it possible to more accurately constrain this ratio by providing rest-frame-optical spectra (which are required -- but not necessarily sufficient -- to accurately correct for dust attenuation; \citealt{Reddy2015}) for thousands of high-redshift galaxies. A relevant ongoing survey is the MOSFIRE Deep Evolution Field (MOSDEF, \citealt{2015ApJS..218...15K}) survey. Using data from this survey and ancillary 3D-HST UV photometry \citep{Skelton2014}, the SFR--$M_*$ relation can be derived separately using both H$\alpha$ and UV fluxes for the same galaxies \citep{2015arXiv150703017S}. A limitation of such observations is that they mostly constrain massive galaxies ($M_*>10^9\msun$), unlike local observations, which provide constraints down to $M_*\simeq 10^6 \msun$, where the effect of supernova feedback -- and thus scatter in the H$\alpha$-to-FUV flux ratio -- is predicted to be much stronger because of the shallower potentials of less-massive galaxies.

We find it worth mentioning that we have reproduced both an increased scatter and a decline in the average H$\alpha$/FUV at low stellar masses. In our simulations, these effects are caused mostly by bursty star formation without accounting for IMF variations \citep[as done in ][]{2003ApJ...598.1076K,2005ApJ...625..754W} or stochastic IMF sampling. Our simulations thus agree with \citet{2012ApJ...744...44W} which suggested that bursty star formation can account for the behaviour of H$\alpha$/FUV in low-mass galaxies.

\subsection{Bursty star formation and the scatter in the SFR--$M_*$ relation}

The SFR--$M_*$ relation plays a central role in galaxy evolution phenomenology, in which the emerging picture is that galaxies build up most of their stellar mass while they are on this relation \citep[e.g.][]{2013ApJ...772..119L} and are fuelled by continuous gas supply \citep{2005MNRAS.363....2K}. According to the common lore, when a merger occurs, galaxies enter the `starburst mode', and this is often believed to be followed by a quenching event in which the star-forming gas not consumed in the starburst is ejected from the galaxy. In many simple analytical models, all galaxies with a given stellar mass are assumed to have the same SFR \citep{2015MNRAS.452.1184M}, whereas in semi-analytical models, scatter in the SFR at fixed $M_*$ is ensured by accounting for the merger and accretion histories of different halos \citep{2015MNRAS.451.2663H}. In large-volume cosmological simulations, the gas flows in galaxies are accounted for, and the SFR varies on timescales of hundreds Myr \citep{2014arXiv1409.0009S}. In these three types of galaxy formation models, galaxies evolve in a quasi-equilibrium state in which the SFR fluctuates slowly with a variability timescale of $\gtrsim$ 100 Myr.

The behaviour of the galaxies in the FIRE simulations challenges this picture. In these simulations, stars are often formed in burst cycles, and it is not unusual that the SFR changes by an order of magnitude or more within a 200-Myr time interval. At $z=0$, this bursty star formation mode is most evident at low masses ($M_*<10^{9}\msun$), whereas at higher masses, a more steadily star-forming mode is present. In the bursty mode, galaxies quickly change from being in a burst to a post-burst phase. When using e.g. a 10-Myr-averaged SFR indicator, one will observe the short-timescale variability of these star formation cycles, but when using an SFR indicator that is sensitive to $\ga 100$ Myr timescales, one will get the impression that the galaxies are in a quasi-equilibrium with a slowly varying SFR. The observation of a tight SFR--$M_*$ relation when using long-timescale SFR indicators can be considered a consequence of the central limit theorem, from which one would expect a tight relation if galaxies are affected by many processes that act on timescales shorter than that to which the SFR indicator employed is sensitive \citep{2014arXiv1406.5191K}.

\subsection{Limited galaxy number statistics}

An important issue to keep in mind when comparing observations with the FIRE simulations is selection effects. The simulations presented in this paper
are all zoom simulations of the environments around a few halos. Properties such as the scatter in the SFR--$M_*$ relation and the scatter in the
SFR(H$\alpha$)$/$SFR(FUV) ratio might therefore be biased by our selection of galaxies from environments that statistically differ from a
cosmologically representative volume. The effect is expected to be most pronounced at $z=0$, where our sample of galaxies comes from only four
different zoom simulations. It is therefore possible that larger samples of simulations would alter the distribution of SFR(H$\alpha$)$/$SFR(FUV)
ratios at $z=0$, where there is some tension between our simulated samples of galaxies and observations.

\subsection{What about galaxy mergers?}

Historically, galaxy mergers and starbursts have been closely connected concepts. Early idealised simulations of galaxy mergers indicated that the mutual tidal forces induced by the interaction could cause otherwise stable discs to develop bars, which subsequently drove strong gas inflows into the central regions of the galaxies \citep[e.g.][]{NegroponteWhite1983,Hernquist1989,MihosHernquist1994,MihosHernquist1996, BarnesHernquist1996,2016arXiv160408205S}. Consequently, the SFR of the system was enhanced considerably: this enhancement could be as much as two orders of magnitude for a short ($\la 100$ Myr) time near final coalescence \citep[e.g.][]{BarnesHernquist1991,MihosHernquist1994,MihosHernquist1996}. Because the simulations indicated that minor mergers could also drive strong starbursts \citep[e.g.][]{MihosHernquist1994minor}, a reasonable conclusion was that the majority of disc galaxies have experienced one or more merger-driven starbursts.

In Section \ref{SFMSBurst}, we noted that the starbursts studied in this work are generally \emph{not} merger-driven but rather a consequence of a combination of clustered star formation and strong stellar feedback. Nevertheless, the SFR enhancements in starbursts exhibited by the FIRE galaxies are comparable to those observed in simulations of merger-induced starbursts (see e.g. Figure \ref{Fig321_MassiveFIRE_1Myr}).

Still, the results presented herein do not rule out that mergers drive strong starbursts. Rather, they indicate that except for massive ($M_* \ga 10^{10} \msun$) galaxies at low redshift ($z \la 1$), galaxies evolve in a quasi-equilibrium characterised by strong bursts of star formation and subsequent periods of `quiescence', \emph{even if they are not actively undergoing mergers}. However, mergers may drive additional burstiness, even for a small subset of the simulated galaxies analysed in this work.

\section{Conclusions} \label{sec:conclusions}
 
In this paper, we studied the short-timescale variability of the SFR in the FIRE simulations by comparing the SFRs calculated using different indicators with different sensitivity timescales. Our analysis compares the SFR averaged over 10- and 200-Myr time intervals, and to compare directly to observations, we also calculated the (unattenuated) H$\alpha$- and FUV-derived SFRs of our simulated galaxies. Our main results are the following:

\begin{itemize}

\item The scatter in the SFR--$M_*$ relation is sensitive to the burstiness of star formation histories. When using H$\alpha$- and FUV-based SFR indicators, the scatter at $z=2$ is 0.39 dex and 0.35 dex, respectively (for a stellar mass cutoff of $M_*> 10^{9.5}\msun$ and  SFR(H$\alpha$)$>2 \msun$ yr$^{-1}$). The scatter is larger for the H$\alpha$-derived SFR because it is more sensitive to short bursts than the FUV-based indicator. We conclude that the difference in H$\alpha$ and FUV scatter is consistent with observations. We note that a direct comparison with observations is complicated by observational uncertainties in deriving the SFR(H$\alpha$), the effect of stochastic IMF sampling, dust reddening and sample selection effects in our simulations.

\item For low-mass simulated galaxies ($M_*<10^{9.5}\msun$), the SFR varies so rapidly that the 10-Myr-averaged SFR can vary by an order of magnitude during a 200-Myr time interval. This result indicates that such galaxies are not evolving steadily on a
`star-forming main sequence'; instead, they have rapidly fluctuating SFRs.

\item The majority of the FIRE galaxies from our sample at $z=0$ have H$\alpha$/FUV ratios consistent with observations. A non-negligible fraction of the simulated galaxies do, however, have too low ratios relative to the observations, indicating that they are in a strong post-burst epoch. This suggests that a small but significant fraction of low-mass galaxies in FIRE have lower SSFR values (i.e. H$\alpha$ equivalent widths) than observed for local-Universe galaxies. This conclusion is independent of whether we treat the effect of stochastic IMF sampling when calculating the H$\alpha$ and FUV fluxes of the simulated galaxies. Accounting for ionizing photons from binaries or resolving further down the GMC mass function may alleviate this tension.

\end{itemize}
We have shown that the amount of burstiness in galaxies can be constrained by comparing with the H$\alpha$ and FUV derived SFR--$M_*$ relation and H$\alpha$/FUV ratios of individual galaxies. We suggest future simulations to take these constraints into account when calibrating feedback models.

\section*{Acknowledgements}
We thank Dan Weisz, Mark Krumholz, Chuck Steidel and the referee for useful discussions. MS thanks the Sapere Aude fellowship program and acknowledges the hospitality of the California Institute of Technology. CCH is grateful to the Gordon and Betty Moore Foundation for financial support. CAFG was supported by NSF through grants AST-1412836 and AST-1517491, by NASA through grant NNX15AB22G, and by STScI through grants HST-AR- 14293.001-A and HST-GO-14268.022-A. This work was supported in part by National Science Foundation Grant No. PHYS-1066293 and the hospitality of the Aspen Center for Physics. DK was supported in part by NSF grant AST-1412153 and funds from the University of California San Diego and the Cottrell Scholar Award. We also acknowledge the following computer time allocations: TG-AST120025 (PI: DK), TG-AST130039 (PI: PH), TG-AST1140023 (PI: CAFG).

\def\aj{AJ}
\def\araa{ARA\&A}
\def\apj{ApJ}
\def\apjl{ApJ}
\def\apjs{ApJS}
\def\apss{Ap\&SS}
\def\aap{A\&A}
\def\aapr{A\&A~Rev.}
\def\aaps{A\&AS}
\def\mnras{MNRAS}
\def\nat{Nature}
\def\pasp{PASP}
\def\aplett{Astrophys.~Lett.}
\def\physrep{Physical Reviews}
\def\nar{New A Rev.}
\def\na{New Astronomy}

\footnotesize{
\bibliographystyle{mn2e}
\bibliography{ref}
}

\clearpage
\appendix

\appendix

\section{SFR(H$\alpha)/$SFR(FUV) at $\lowercase{z}=2$}\label{HalphaFUVRedshift2}

\begin{figure}
\centering
\includegraphics[width = 0.48 \textwidth]{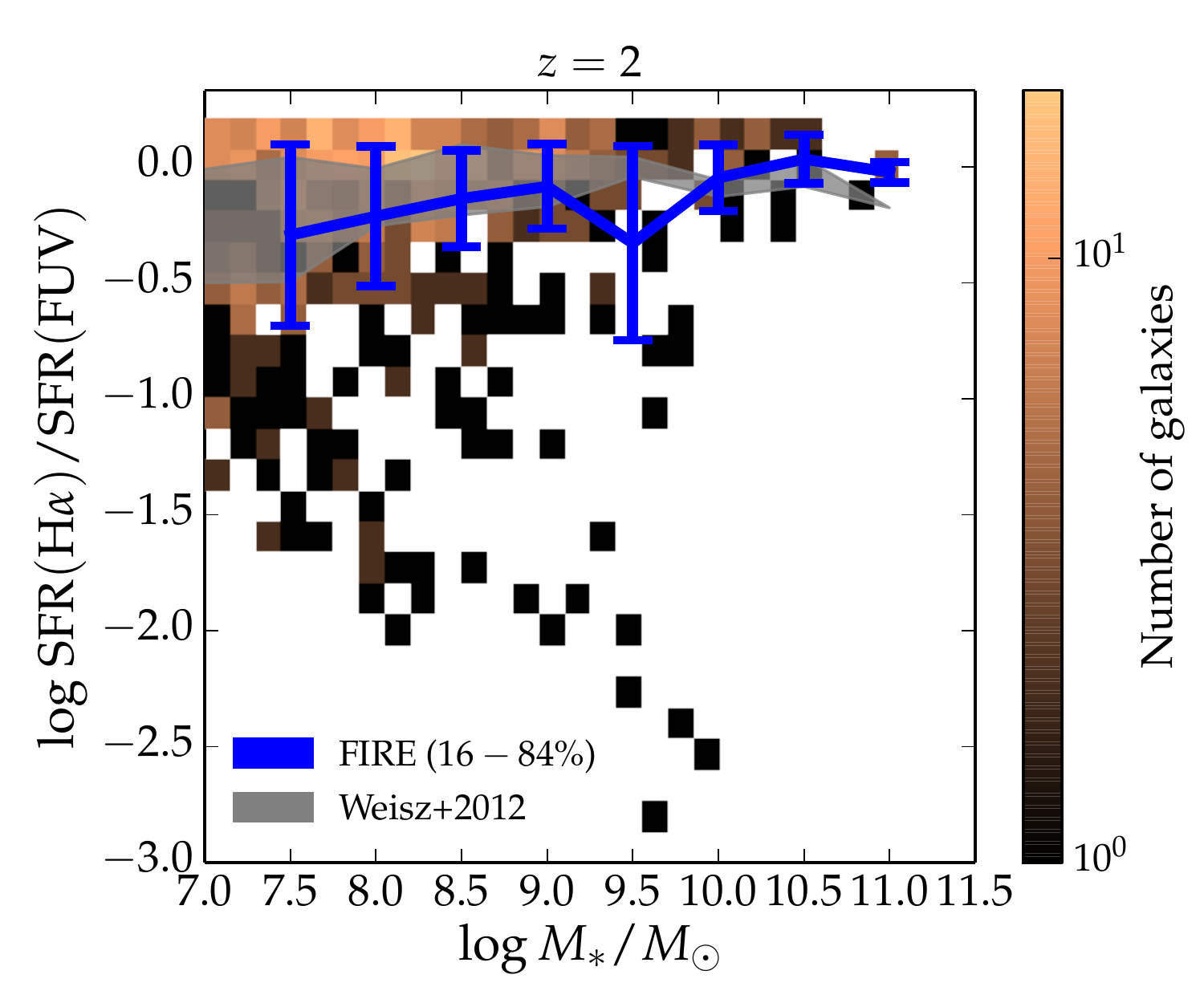}
\caption{SFR(H$\alpha)/$SFR(FUV) versus stellar mass for the simulated galaxies at $z=2$ (\emph{background histogram}). The \emph{blue line} and error bars indicate the median values and 16--84th percentile ranges, respectively, for different mass bins. This figure is similar to Figure~\ref{Fig08_Weisz_SLUG}, where we studied the $z=0$ sample. Because there are no relevant observations to which we can compare the $z=2$ simulations, we compare with the $z=0$ observations from \citet{2012ApJ...744...44W}. Compared with the simulated galaxies at $z=0$ (Figure~\ref{Fig08_Weisz_SLUG}), we see that at $z=2$, high-mass galaxies are burstier than at $z = 0$, and  there is a larger number of galaxies in very strong post-burst states, with SFR(H$\alpha)/$SFR(FUV)$\lesssim 0.01$.}
\label{Fig1023_Weisz2_Redshift2}
\end{figure}

In Section~\ref{ObservationComparison}, we used the mass dependence of the SFR(H$\alpha)/$SFR(FUV) ratio of galaxies to constrain the amount of burstiness in our simulations at low redshift. The behaviour of this ratio at $z=2$ is shown in Figure~\ref{Fig1023_Weisz2_Redshift2}. No observations can directly constrain SFR(H$\alpha)/$SFR(FUV) at $z=2$, so we again compare to the $z=0$ observations from \citet{2012ApJ...744...44W}. There is a trend that the SFR variability in FIRE decreases with increasing stellar mass. An exception is the mass bin at $M_*=10^{9.5}\msun$, which features a higher fraction of strong post-burst galaxies than any other mass bin. This is caused by a handful of extreme events with SFR(H$\alpha)/$SFR(FUV)$\lesssim 0.01$. Whether this is a genuine physical effect or an artifact of small-number statistics is unclear.

A relevant effect worth highlighting is that massive galaxies with $M_*\gtrsim 10^{10}\msun$ exhibit a larger scatter in their SFR(H$\alpha)/$SFR(FUV) ratios than simulated $z=0$ galaxies of the same mass (see Figure~\ref{Fig08_Weisz_SLUG}). This is consistent with other parts of analysis that revealed massive galaxies to be more bursty at $z=2$ than at $z=0$ (e.g. Figure~\ref{Fig06A_BurstSFRFrac}).

The intervals containing the 16-84$\%$ percentiles of the SFR(H$\alpha)/$SFR(FUV) distributions at $z=2$ are remarkably similar to the $z=0$ versions. The biggest differences are that the massive galaxies have a larger spread at high redshift than at low redshift and that a few extreme galaxies with SFR(H$\alpha)/$SFR(FUV)$\lesssim 0.01$ increase the width of the interval around $M_*=10^{9.5}\msun$ at $z=2$.

\label{lastpage}

\end{document}

%% file: JournalAbbr.tex
\def\aj{AJ}%
\def\araa{ARA\&A}%
\def\apj{ApJ}%
\def\apjl{ApJ}%
\def\apjs{ApJS}%
\def\apss{Ap\&SS}%
\def\aap{A\&A}%
\def\aapr{A\&A~Rev.}%
\def\aaps{A\&AS}%
\def\mnras{MNRAS}%
\def\pasp{PASP}%
\def\nat{Nature}%
\def\aplett{Astrophys.~Lett.}%
\def\na{New Astronomy}%
\def\physrep{Phys.~Rep.}%